\documentclass[journal]{IEEEtran}
\IEEEoverridecommandlockouts
\usepackage{amssymb}
\usepackage{amsthm}
\usepackage{multirow}
\usepackage{inputenc}
\usepackage{enumerate} 
\usepackage{textcomp}
\usepackage{listings}
\usepackage{array}
\usepackage{tikz}
\usetikzlibrary{shapes.geometric, arrows}
\usepackage{dirtytalk}
\usepackage[switch,mathlines]{lineno}
\usepackage{soul}
\usepackage{xfrac}
\usepackage{bbm}
\usepackage{breqn}
\usepackage{graphicx}
\usepackage{cite}
\newtheorem{theorem}{Theorem}
\usepackage{algorithm,algpseudocode}
\algnewcommand{\Inputs}[1]{%
  \State \textbf{Inputs:}
  \Statex \hspace*{\algorithmicindent}\parbox[t]{.8\linewidth}{\raggedright #1}
}
\algnewcommand{\Initialize}[1]{%
  \State \textbf{Initialize:}
  \Statex \hspace*{\algorithmicindent}\parbox[t]{.8\linewidth}{\raggedright #1}
}
\usepackage{algpseudocode}

\usepackage{cancel}
\usepackage{cite}
\usepackage{bm}
\usepackage{subfig}
\usepackage{float}

\begin{document}
\title{Measurement Uncertainty Impact on Koopman Operator Estimation of Power System Dynamics \author{Pooja Algikar,~\IEEEmembership{Member,~IEEE}, Pranav Sharma,~\IEEEmembership{Member,~IEEE}, \\ Marcos Netto,~\IEEEmembership{Senior~Member,~IEEE}, and Lamine Mili,~\IEEEmembership{Life Fellow,~IEEE}}
\thanks{This work was authored in part by the National Renewable Energy Laboratory, operated by Alliance for Sustainable Energy, LLC, for the U.S. Department of Energy (DOE) under Contract No. DE-AC36-08GO28308. Funding provided by U.S. Department of Energy Office of Electricity. The views expressed in the article do not necessarily represent the views of the DOE or the U.S. Government. The U.S. Government retains and the publisher, by accepting the article for publication, acknowledges that the U.S. Government retains a nonexclusive, paid-up, irrevocable, worldwide license to publish or reproduce the published form of this work or allow others to do so, for U.S. Government purposes.}}
\maketitle
\begin{abstract} 
Sensor measurements are mission critical for monitoring and controlling power systems because they provide real-time insight into the grid operating condition; however, confidence in these insights depends greatly on the quality of the sensor data. Uncertainty in sensor measurements is an intrinsic aspect of the measurement process. In this paper, we develop an analytical method to quantify the impact of measurement uncertainties in numerical methods that employ the Koopman operator to identify nonlinear dynamics based on recorded data. In particular, we quantify the confidence interval of each element in the push-forward matrix from which a subset of the Koopman operator's discrete spectrum is estimated. We provide a detailed numerical analysis of the developed method applied to numerical simulations and field data collected from experiments conducted in a megawatt-scale facility at the National Renewable Energy Laboratory.
\end{abstract}
\section{Introduction}

\IEEEPARstart{T}{he} electric power grids are networked dynamical systems with thousands of components that exhibit dynamic behavior. The real-time operation of these systems, subject to economic and reliability constraints, encompasses numerous challenges; see \cite{Wollenberg2012} for an overview. In the pursuit of clean energy, these challenges are compounded by integrating into the grid inverter-based resources \cite{Matevosyan2021}, whose dynamics are driven by controls and fundamentally differ from legacy synchronously rotating machinery. Inverter-based resources are exacerbating the nonlinear dynamics of power systems \cite{fan2022data}.

On the other hand, advancements in sensing, communications, data analysis, and computing power have propelled the incorporation of data-driven methods into grid operations \cite{tan2016survey}. In particular, high-fidelity synchrophasor measurements and sampled values \cite{IEC61869-9} are suitable for characterizing power system dynamics; this characterization is essential for the resilient and stable operation of future electric power grids \cite{brahma2016real, farantatos2009pmu}. As a result, a wealth of data-driven methods to characterize the grid dynamics have emerged or resurfaced with renewed interest. Examples of legacy methods include Prony analysis \cite{Netto2018}, the matrix pencil method \cite{Crow2005}, and the subspace identification method \cite{Zhou2006}. However, all these methods assume that the underlying dynamics are linear. Hence, there is a limit to the dynamic phenomena they can capture. On the contrary, data-driven methods derived from the Koopman operator formalism \cite{Mauroy2020, Brunton2022} can reveal linear and nonlinear dynamics and have gained momentum among researchers on power systems. Applications include coherency identification \cite{susuki2011nonlinear}, stability analysis \cite{susuki2013nonlinear}, state \cite{Netto2018AEstimation} and parameter \cite{susuki2018estimation} estimation, selective modal analysis \cite{Netto2019, Sharma2022Data-DrivenMeasurements}, and model identification \cite{Sinha2020DataDynamics}. This paper focuses on numerical methods that are based on the Koopman operator. 


Dynamic mode decomposition (DMD) \cite{Schmid2010} is a class of numerical methods with connections to the Koopman operator formalism \cite{Rowley2009} widely used to perform data-driven analysis of nonlinear dynamical systems \cite{Barocio2015}. It is well-known that the DMD method is sensitive to data imperfections. Duke \emph{et al.} \cite{duke2012error} explore the impact of factors such as the quantity and quality of the data, the signal-to-noise ratio, and the quantity of the ensemble on the accuracy of DMD. Dawson \emph{et al.} \cite{Dawson2014CharacterizingDecomposition} quantify the bias in the estimation of the Koopman operator by the DMD method resulting from additive noise levels in the measurements and propose corrective measures. 
N{\"u}ske \emph{et al.} \cite{nuske2023finite} derive probabilistic bounds on how well the extended DMD algorithm \cite{Williams2015ADecomposition} captures the underlying system dynamics. They consider both ergodic and independent identically distributed (i.i.d.) samples for systems described by ordinary and stochastic differential equations. They quantify the approximation error using the Frobenius norm between the Koopman operator estimated from finite samples and the Galerkin projection of the Koopman generator onto the space of observables. In formulating theoretical bounds on error, the authors decompose the variance of the random matrix that constitutes the basis functions into terms related to the asymptotic contribution and the impact of the number of data points.
Zhang and Zuazua \cite{zhang2023quantitative} analyze the convergence of a variant of the generator-extended DMD algorithm. The authors establish bounds for the approximation errors but do not consider measurement noise. 
Based on the convergence of the coefficients of the approximate stiffness matrix of the dictionary functions to a normal probability distribution, the authors arrive at the convergence of the Koopman operator in the matrix norm in terms of the largest and smallest eigenvalues of the stiffness matrix.

Diverse methods have been introduced to deal with uncertainties beyond measurement errors in the context of the Koopman operator. Lian and Jones \cite{Lian2020OnOperators} propose a probabilistic framework based on Gaussian processes for propagating model uncertainty through the observables space. Xu \emph{et al.} \cite{Xu2022PropagatingModel} propagate parametric uncertainty using a Koopman operator-based surrogate model of a dynamical system. Additionally, \cite{Matavalam2022PropagatingOperators} present the Koopman operator-based moment propagation to estimate the upper and lower bounds of the states due to uncertain initial conditions. The landscape of the current literature reveals a conspicuous gap regarding quantifying measurement uncertainty in the data-driven estimation of the Koopman operator. Recent studies prioritize on determining overall bounds for approximating the Koopman operator in the Frobenius norm over detailed element-wise uncertainty. When applying the Koopman operator to quantify uncertainty in power system dynamics, extensive efforts have focused on assessing uncertainty originating from diverse sources within the states of the underlying dynamic process. No other work has been presented in the literature to quantify the uncertainty associated with the framework of data-driven estimation algorithms of the Koopman operator except that presented in \cite{Lian2020OnOperators}. 

Our approach focuses on empirically quantifying errors in the data-driven approximation of the Koopman operator, diverging from the methods described in \cite{nuske2023finite, zhang2023quantitative}. These errors are attributed to the measurement process in recording the finite data of the states. Specifically, first the measurement uncertainty is represented by sampling random data centered around these recorded data points in i.i.d. fashion with variances of the states. The uncertainty expressed in terms of these variances is then propagated through the algorithmic steps of EDMD and DMD. The variance considered for the i.i.d. sampling is obtained either by the computation of variances of the states measured over steady-state period or from the manufacturers' specifications. We derive analytical expressions for the first and second moments of the elements of the Koopman operator by leveraging principles rooted in random matrix theory. These expressions are based on the corresponding moments of the states obtained from sensor measurements' characteristics. Eventually, the quantitative uncertainty bounds of the individual elements of the Koopman operator are given by variance. Additionally, we provide expressions for the probability density functions of the probability distributions of both the Koopman eigenfunctions and Koopman modes. In contrast to the method presented in \cite{Dawson2014CharacterizingDecomposition}, our approach primarily concentrates on the second moments, leveraging the randomness exhibited by the Koopman operator when estimated using the non-robust EDMD and DMD algorithms.

This paper is organized as follows. Section \ref{sec2} gives some background on the theory of the Koopman operator followed by its estimation algorithms purely based on the observed data. Section \ref{sec3} presents the development of the proposed algorithm, measurement uncertainty quantification (MUQ). Section \ref{sec4} illustrates the implementation of the proposed MUQ for the application of a power system through an example of 3 buses. In Section \ref{section_realdata} we deploy the proposed framework on real system data and present a detailed analysis of the impact of PMU measurement uncertainties on Koopman operator-based dynamic characterization. Section \ref{sec5} concludes the paper and outlines future work.

\section{Background}\label{sec2}
\subsection{Koopman Operator Theory}
Let us define the following dynamical system: 
        \begin{equation}\label{eq1}
            f(\mathbf{x})=\dot{\mathbf{x}}.
        \end{equation}
Here, we assume that the state space $\mathbf{x}\in\mathbb{S}$ of the dynamical system forms a linear vector space with certain geometrical properties and that the rule of evolution has some degree of regularity, which is denoted as $f(\cdot)$.
Let us assume that the solution to \eqref{eq1} exists. Let us define the flow map, $F^{t}(\mathbf{x})$, which takes $\mathbf{x}\in\mathbb{R}^{n}\in  \mathbb{S}$ from its initial state, $\mathbf{x}_{0}$, to the state $\mathbf{x}^{t}$ at time $t$ given by
\begin{equation}
     F^t(\mathbf{x}_{0}) = \mathbf{x}_{0} + \int_{\mathbf{x}=\mathbf{x}_{0},t'=t_{0}}^{t'=t} f(\mathbf{x}(t_0)) \, dt'.
\end{equation}

The Koopman operator is a composition operator. 
The Koopman operator lifts the dynamics of the system from a finite-dimensional nonlinear state space to an infinite-dimensional observable linear space. Consequently, we project the dynamics from the state space $\mathbb{S}$ by $g(F^{t}(\mathbf{x}))$ to an infinite observable space to get the linear rule of evolution defined by the Koopman operator, $\mathbf{U}^{t}.$
Formally, we have an expression for the evolution of the system from time $t_{0}$ to $t$ given by:
\begin{equation}
    g(\mathbf{x}_{t})=g(F^{t}(\mathbf{x}_{0}))=\mathbf{U}^{t}g(\mathbf{x}_{0}).
\end{equation}
The eigenfunctions, $\phi_{j}$, associated with the $j^{th}$ eigenvalue, ${\lambda}_{j}$, of the Koopman operator, $\mathbf{U}^{t}$, are defined as:
    \begin{equation}
        \mathbf{U}^{t}\phi_{j}=e^{{\lambda}_{j}t}\phi_{j}.
    \end{equation}

Now, we assume that all the observables lie in the linear span of the Koopman eigenfunctions, which are the right eigenvectors of the Koopman operator; therefore, the observables can be constructed as a linear combination of eigenfunctions given by
    \begin{equation}
        g(\mathbf{x})=\sum_{k=0}^{\infty}c_{k}\phi_{k}(\mathbf{x}),
    \end{equation}
yielding
$\mathbf{U}^{t}g(\mathbf{x})=\sum_{k=0}^{\infty}c_{k}e^{{\lambda}_{k}t}\phi_{k}(\mathbf{x})$. 
The coefficients $c_{k}$, of these linear expansions are called Koopman modes, which inform us of the shape or structure within the data that evolves with the eigenvalues of the Koopman operator, ${\lambda}_{k}$. Note that the principal eigenfunctions are not all the right eigenvectors of the Koopman operator but only those that are linear combinations of observables. 

\subsection{Data-Driven Estimation of Koopman Operator}
Let us assume that we have access to a set of time series measurement vectors, $\bm{g}=[g^{1},\hdots,g^{{T}}]^{'}$, which are assumed to be functions of the state vector, $\mathbf{x}$. Estimation of the Koopman operator from a set of $m$ measurements of observables given by $\mathbf{g}$ is called a data-driven estimation of the Koopman operator. Two popular methods proposed in the literature are DMD and EDMD. 

\subsubsection{Dynamic Mode Decomposition}
The Koopman operator represents a transformation that maps an observable function of the state of the system in the previous time steps, $\mathbf{X}=[\mathbf{x}_{1},\hdots,\mathbf{x}_{{T-1}}]^{'}$, to its value at the next time steps, $\mathbf{Y}=[\mathbf{x}_{2},\hdots,\mathbf{x}_{T}]^{'}$. It is usually estimated based on the DMD, which is based on the Moore-Penrose pseudo-inverse given by
\begin{equation}
    \mathbf{K}=\mathbf{X}^{\dag}\mathbf{Y}.
\end{equation}
\subsubsection{Extended Dynamic Mode Decomposition}
It does not require us to store the observations in all time steps $m=T-1$. Instead, using the vector-valued observable dictionary or basis functions, $\bm{\psi}:\mathbb{R}^{n}\to\mathbb{R}^{N_{k}}$, the Koopman operator is estimated as: 
\begin{equation}
    \mathbf{K}=\mathbf{G}^{\dag}\mathbf{A},
\end{equation}
where $\mathbf{G}=\sum_{i=1}^{T} \bm{\psi}(\mathbf{x}_{i})\bm{\psi}(\mathbf{x}_{i})^{T}$ and $\mathbf{A}=\sum_{i=1}^{T} \bm{\psi}(\mathbf{x}_{i})\bm{\psi}(\mathbf{y}_{i})^{T}$. Thus, developed data-driven estimation of the Koopman operator paves the way for characterization of nonlinear dynamics of an underlying system using measurements. In recent years, various applications have been developed for system stability, parameter identification, trajectory prediction, and coherency identification for modern power systems using the Koopman theoretical framework. In the next section, we will illustrate the impact of measurement uncertainty on the Koopman operator-based system identification applied to power systems.   
\section{Measurement Uncertainty Quantification}\label{sec3}
Let us gather a single observation of the state vector, $\mathbf{x}\in \mathbb{R}^{n}$, for the time period $1: {T-1}$ in $\mathbf{X}_{obs}\in \mathbb{R}^{m\times n}$ and the observations for the time period $2: {T}$ in $\mathbf{Y}_{obs}\in \mathbb{R}^{m\times n}$. 
Let us denote the Koopman operator, linearly approximating the nonlinear dynamical system, estimated using DMD as $\hat{\mathcal{K}}=\mathbf{X}_{obs}^{\dag}\mathbf{Y}_{obs}$. The measurement uncertainty problem is formulated as 
\begin{equation}
    \hat{\mathcal{K}}=\mathcal{K}_{0}+\bm{E}, 
\end{equation}
where $\bm{E}\in\mathbb{R}^{m\times n}$ denotes the true error term.
We adopt a random matrix theory approach to perform MUQ characterized by the variance of the error term, Var$(\bm{E})$. In this framework, our data set consists of observations for $n$ states over a period of time, $T$ gathered in $\mathcal{D}_{obs}=\{\mathbf{X}_{obs},\mathbf{Y}_{obs}\}$. Each element of the input and output data matrix, denoted as ${X}_{ij}$ and $Y_{ij}$, respectively, is treated as a random variable. We assume that each random variable follows a Gaussian distribution, characterized by a mean equal to the observed quantity, $X_{{obs}_{ij}}$, and a standard deviation, denoted as $\sigma_{ij}$. Similarly, ${Y}_{ij}\sim \mathcal{N}(Y_{{obs}_{ij}},\sigma_{ij}^{2})$.

\subsection{Characterization of Measurement Uncertainty}
Uncertainty is often expressed using statistical metrics, such as standard deviations, variances, confidence intervals, and probability distributions. These metrics provide a quantitative measure of the expected deviation between a measured value and the true value. We characterize the measurement uncertainty from the period of observations of states in which they have remained steady. As each state is consistently measured using the same device throughout the entire time period, $T$, we can characterize the measurement uncertainty by calculating the standard deviation of that particular state throughout the duration of $T$;
therefore, each row, $\mathbf{x}_{i}$, of the data matrix, $\mathbf{X}$, follows a multivariate Gaussian distribution, defined as: 
\begin{equation}
    \mathbf{x}_{i}\sim \mathcal{N}(\mathbf{x}_{{obs}_{i}},\bm{\Sigma}),
\end{equation}
where $\bm{\Sigma}=\textrm{diag}(\sigma_{1}^{2},\dots,\sigma_{n}^{2})$ for $i=1,\dots,T-1.$ Similarly, $ \mathbf{y}_{i}\sim \mathcal{N}(\mathbf{y}_{{obs}_{i}},\bm{\Sigma})$ for $i=2,\hdots,T.$
Another way to characterize the measurement uncertainty is to fix the covariance matrix, $\bm{\Sigma}$, to the values obtained from the manufacturer of the measurement devices. 
\begin{figure}[t!]%
\centering
\subfloat[\centering ]{{\includegraphics[height=4.5cm,width=3.7cm]{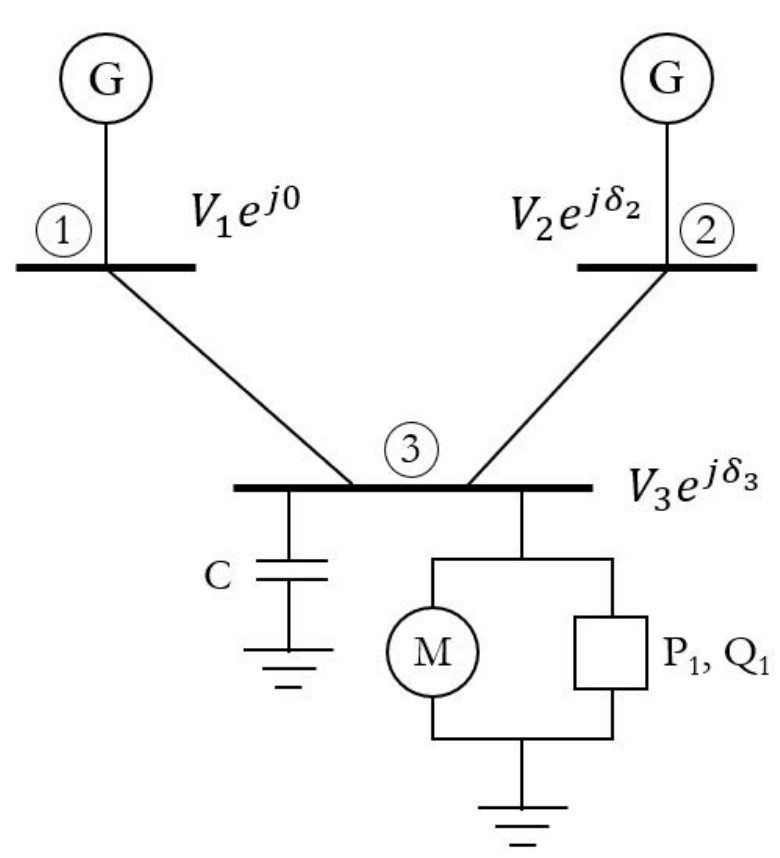}}}%
\quad
\subfloat[\centering  ]{{\includegraphics[height=4.5cm,width=4.8cm]{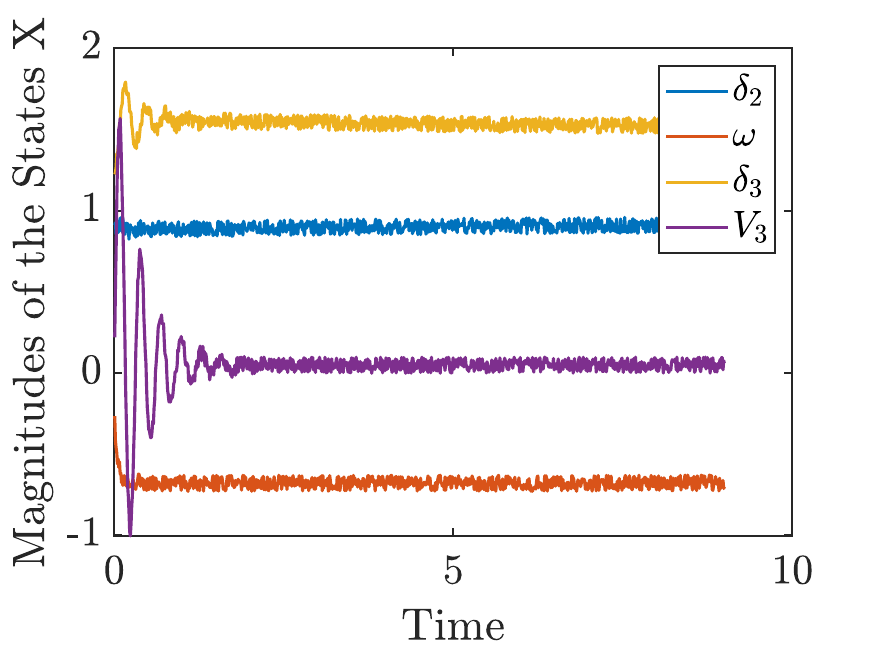}}}%
\caption{(a) 3-bus system with 2 generators and a load and (b) data gathered in $\mathbf{D}$ from the simulation of the 3-bus system.}\label{3_bus_fig}
\end{figure}%
\begin{figure*}[t!]%
\centering
\subfloat[\centering ]{{\includegraphics[height=4cm,width=4.5cm]{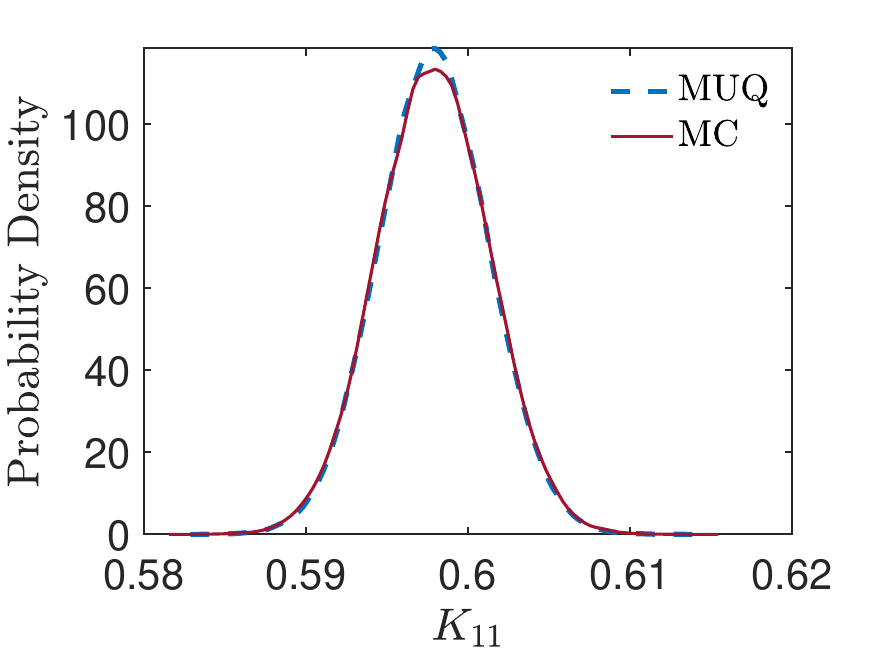}}}%
\subfloat[\centering  ]{{\includegraphics[height=4cm,width=4.5cm]{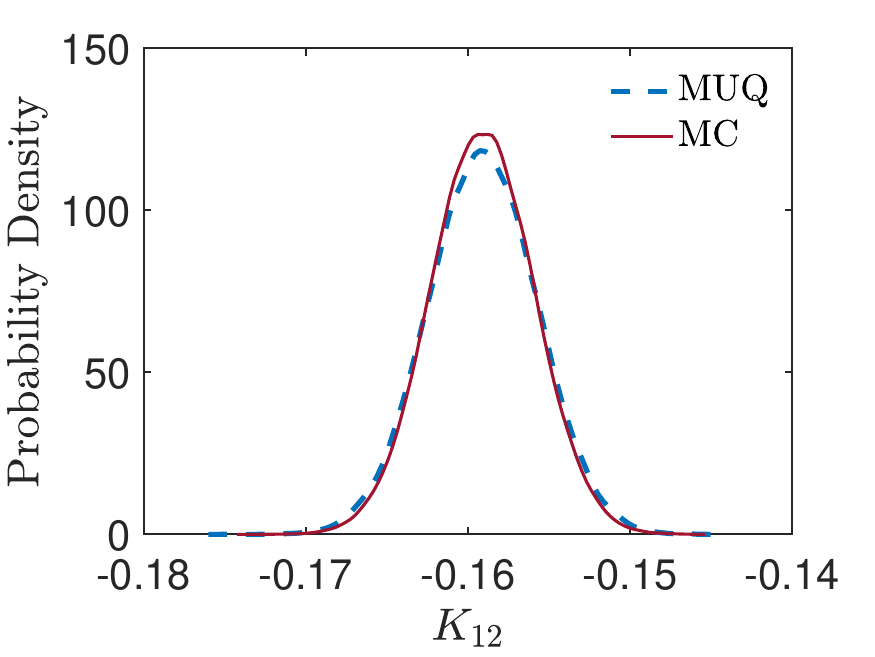}}}%
\subfloat[\centering ]{{\includegraphics[height=4cm,width=4.5cm]{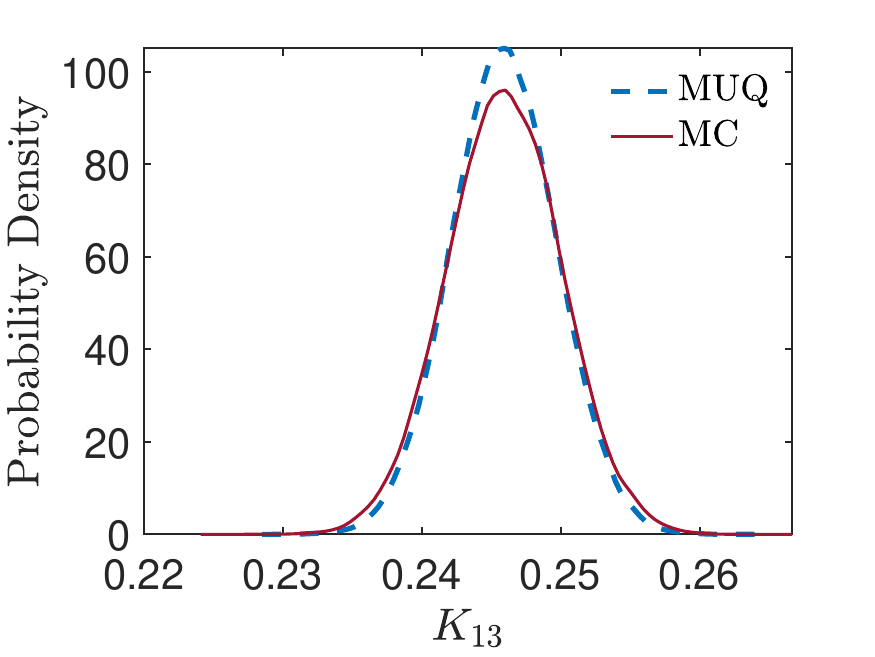}}}%
\subfloat[\centering  ]{{\includegraphics[height=4cm,width=4.5cm]{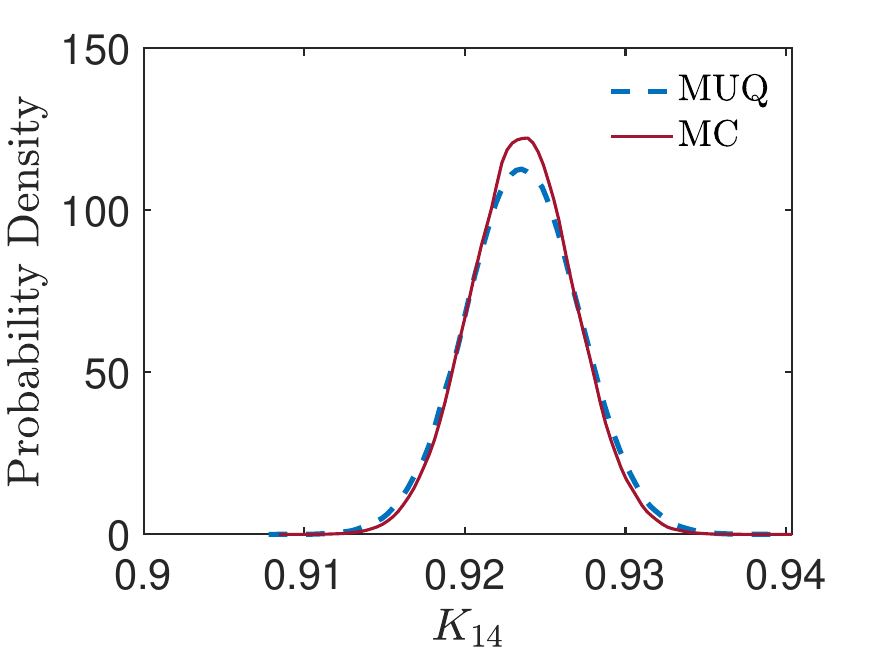}}}%
\;
\subfloat[\centering ]{{\includegraphics[height=4cm,width=4.5cm]{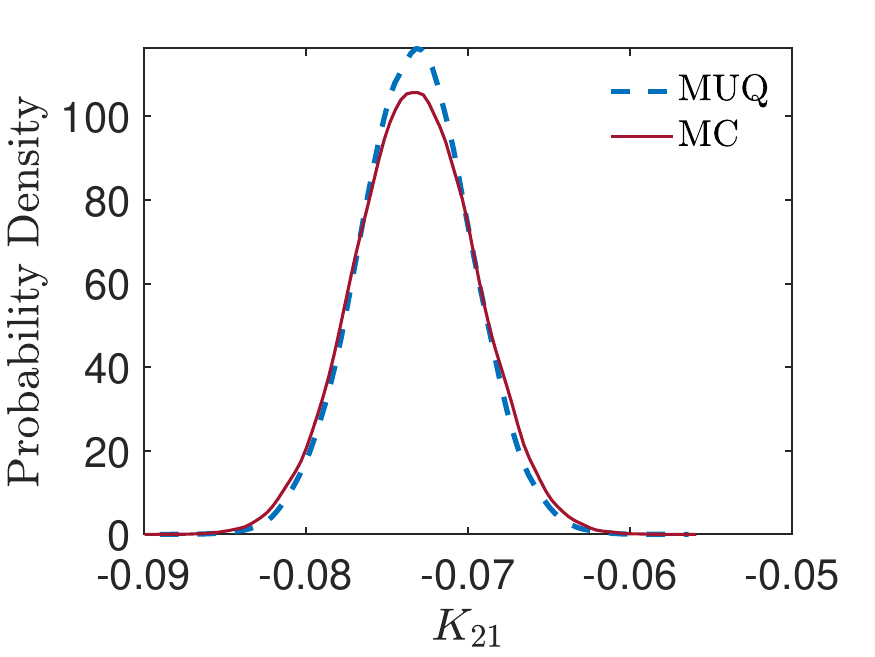} }}%
\subfloat[\centering ]{{\includegraphics[height=4cm,width=4.5cm]{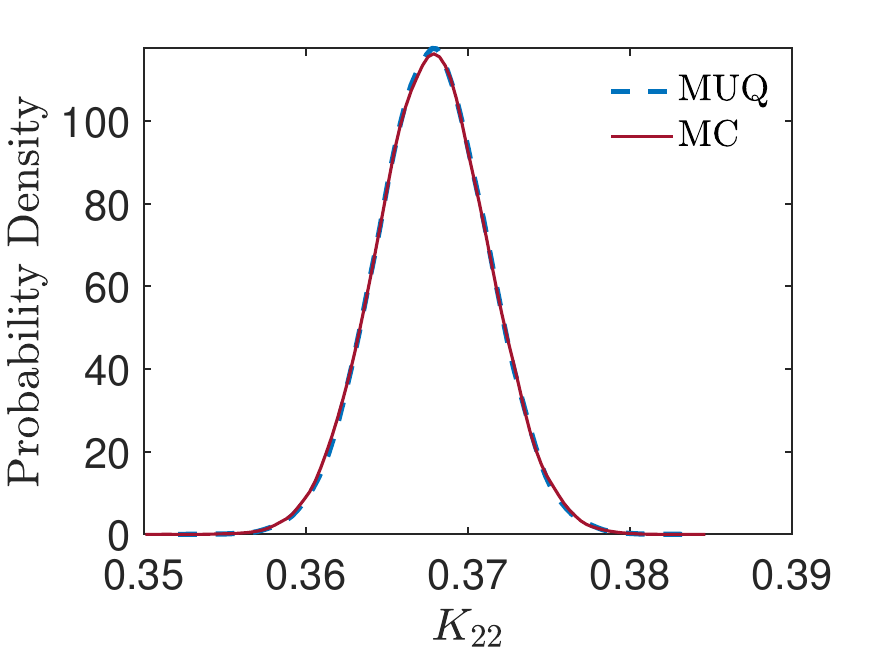} }}%
 \subfloat[\centering ]{{\includegraphics[height=4cm,width=4.5cm]{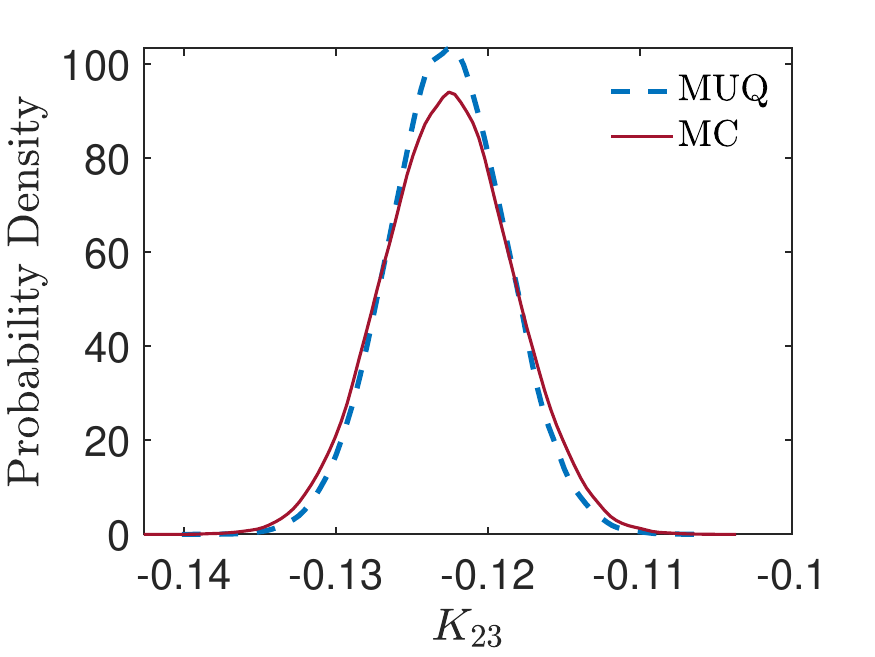}}}%
\subfloat[\centering  ]{{\includegraphics[height=4cm,width=4.5cm]{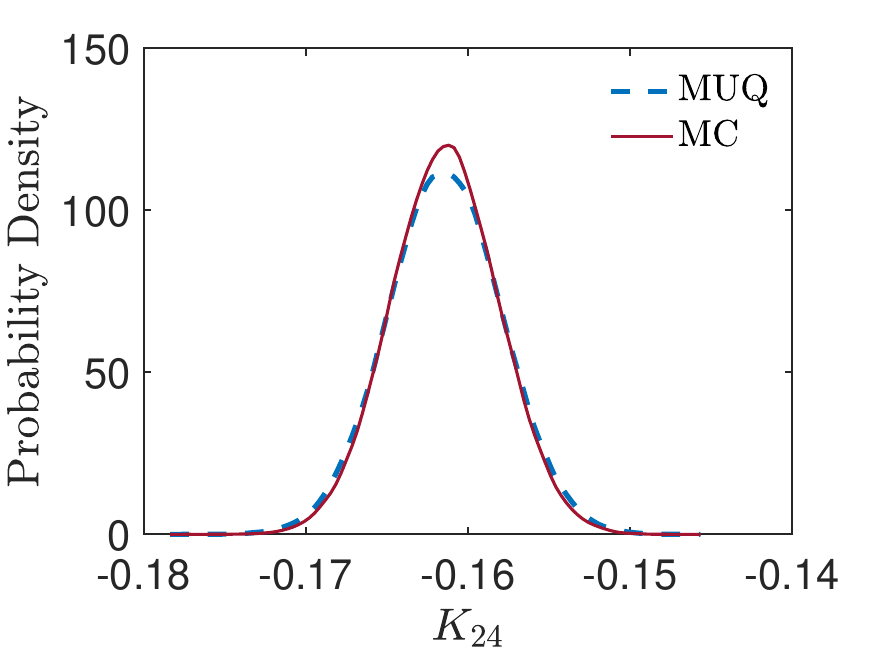} }}%
\caption{{Comparison between the true variance, $R$, and the ones obtained from the proposed MUQ algorithm, $S$, for the elements (a) $K_{11}$, (b) $K_{12}$, (c) $K_{13}$, (d) $K_{14}$, (e) $K_{21}$, (f) $K_{22}$, (g) $K_{23}$, and (h) $K_{24}$ of the Koopman operator, $\mathbf{K}$, estimated via DMD.}}%
\label{prob_1}%
\end{figure*} 
\subsection{MUQ using EDMD Estimation}
Let us consider a quadratic polynomial function as a basis function that maps $n$-dimensional $i^{th}$ input, $\mathbf{x}_{i}$, to $\bm{\psi}(\mathbf{x}_{i})\in\mathbb{R}^{2n}$. 
Let us construct the feature matrix, $\bm{G}=[\bm{\psi}(\mathbf{x}_{1}), \hdots,\bm{\psi}(\mathbf{x}_{m})]\in \mathbb{R}^{m\times 2n}$, involving the quadratic basis functions of the data as:
\begin{equation}
\small
    \bm{G}\in\mathbb{R}^{m\times 2n}=\begin{bmatrix}
    X_{11} & \hdots& X_{1n} & X_{11}^{2} & \hdots &  X_{1n}^{2}\\
    \vdots & \hdots& \vdots & \vdots & \hdots &  \vdots\\
    X_{T1} & \hdots& X_{mn} & X_{m1}^{2} & \hdots &  X_{mn}^{2}\\
\end{bmatrix}.
\end{equation}
\begin{algorithm}[t!]
\small
\caption{MUQ in Koopman Operator via DMD}\label{Al1}
\begin{algorithmic}
\State \textbullet\ Obtain the data $\mathcal{D}_{obs}=\{\mathbf{X}_{obs},\mathbf{Y}_{obs}\}$

\State \textbullet\ Obtain the variance matrix $\bm{\Sigma}=\textrm{diag}(\sigma_{1}^{2},\hdots,\sigma_{n}^{2})$

\State \textbullet\ Generate $N$ random samples around $X_{obs},Y_{obs}$ following the distribution $X_{ij}\sim \mathcal{N}(X_{{obs}_{ij}} ,\sigma_{j}^2)$ and $Y_{ij}\sim \mathcal{N}(Y_{{obs}_{ij}} ,\sigma_{j}^2)$

\State \textbullet\ Initialize $k=1,i=1,j=1$
\For {$i:1:n$}  
   \For {$j:1:n$} 
\State {$\boldsymbol{-}$} $S(i,j)=\frac{\sigma_{j}^{2}}{\sigma_{i}^{2}({m}-n-1)}$;
\If{$i\neq j $}
    \State {$\boldsymbol{-}$}$ S(i,j)= \frac{1}{(m-n-1)}$ 
\EndIf
\State {$\boldsymbol{-}$} $j = j+1 $
\EndFor
\State {$\boldsymbol{-}$} $ i = i+1 $
\EndFor
\State \textbullet\ $\mathbb{E}[K_{ij}]=K_{{obs}_{ij}}$
\State \textbullet\ $\textit{Var}[K_{ij}]=S_{ij}$
\end{algorithmic}
\end{algorithm}
The squared standardized random variables follow a normal distribution, that is,  $X_{ij}^{2}\sim\mathcal{N}(0,3\sigma_{j}^{4})$ and $Y_{ij}^{2}\sim\mathcal{N}(0,3\sigma_{j}^{4})$. 
Accounting for the additional random variable in the feature space, we get the Koopman operator estimation as:
\begin{equation}
   \hat{\mathbf{K}}=\bm{G}^{\dag}\bm{A},
\end{equation}
where $\bm{A}=[\bm{\psi}(\mathbf{y}_{1}), \hdots,\bm{\psi}(\mathbf{y}_{m})].$
The variance matrix characterizing the measurement uncertainty for the $2n-$dimensional feature space is given by $\bm{\Sigma}=\textrm{diag}(\sigma_{1}^2,\hdots,\sigma_{n}^2,3\sigma_{1}^{4},\hdots,3\sigma_{n}^{4})$. 
EDMD fashions the Koopman operator estimation in the least-squares framework as follows: 
        \begin{align}
        \small
            \mathbf{K}&=\mathbf{G}^{\dag}\mathbf{A},\\
            &=\begin{bmatrix}
                G_{1,1} & \hdots & G_{1,2n}\\
                \vdots &   \ddots  &  \vdots\\
                G_{{m},1} & \hdots  & G_{{m},2n}\\
            \end{bmatrix}^{\dag} \begin{bmatrix}
                A_{1,1} & \hdots &  A_{1,2n}\\
                \vdots &   \ddots  &  \vdots\\
                A_{{m},1} & \hdots  & A_{{m},2n}\\
            \end{bmatrix},
        \end{align}
where $\mathbf{G}^{\dag}$ denotes the Moore-Penrose inverse of the matrix $\mathbf{X}.$
EDMD accounts for random errors in the measurements and assigns the same confidence to each measurement in the estimation procedure. 

Now, measurement uncertainty is a broader term that has a probabilistic basis attributed to the spread of the estimated value, $\hat{\mathcal{K}}$, from the true value, $\mathcal{K}_{0}$; therefore, when we refer to the quantification of measurement uncertainty in the context of the Koopman operator, we essentially address the quantification of its algorithmic uncertainty. 
\begin{theorem}\label{th1}
Let the columns of $\bm{X}=[\bm{x}_{1},\hdots,\bm{x}_{m}]'\in \mathbb{R}^{n\times m}$ be a sample, $\bm{x}_{i}$, $i=1,\hdots,T,$, from $\mathcal{N}_{n}(0,\bm{\Sigma})$, $n-$ variate normal probability distribution with a positive definite variance matrix, $\bm{\Sigma}.$ The sum of squares matrices, given by $\bm{S}=\mathbf{X}\mathbf{X}'$, follows a $n$-variate Wishart distribution with $T$ degrees of freedom, denoted by $W_{n}(\bm{\Sigma},m)$. If $T>n+3$, then the first and second moments of $\mathbf{X}^{\dag}$ are given by:
\begin{align}
    \mathbb{E}[X_{ij}^{\dag}]&=0,\\
    \mathbb{E}[X^{{\dag}^{2}}_{ij}]&=\frac{1}{\sigma_{ij}^{2}m(m-n-1)}.
\end{align}
\end{theorem}
\begin{proof}
By invoking Theorem 2.1 from the seminal work by Cook et al. \cite{Cook2011OnMatrix}, we establish the first moment of the generalized inverse, $W^{\dag}$, of a matrix, $W$, following the distribution $W_{n}(\mathbf{I}_{n},{m})$:
\begin{equation}\label{eqth1}
    \mathbb{E}[W^{\dag}]=\frac{n}{m(m-n-1)}\mathbf{I}_{T}
\end{equation}
Considering that $\mathbb{E}[W^{\dag}]=(\mathbf{X}\mathbf{X}')^{\dag}$, we derive: 
\begin{equation}\label{eqth2}
    \mathbb{E}[\mathbf{X}^{'\dag}\mathbf{X}^{\dag}]=\frac{n}{m(m-n-1)}\mathbf{I}_{m}
\end{equation}
Examining the diagonal elements of the mean of this matrix, we observe a direct relationship with the variance of each element:
 \begin{equation}
      \mathbb{E}[\mathbf{X}^{'\dag}\mathbf{X}^{\dag}]=n\mathbb{E}[(X^{{\dag}}_{ij})^{2}]\mathbf{I}_{m}.
 \end{equation}
This relationship leads to the following insightful result:
\begin{equation}
     \mathbb{E}[X^{{\dag^{2}}}_{ij}]=\frac{1}{\sigma_{ij}^{2}{m}({m}-n-1)},
\end{equation}
for a case where $W\sim W_{n}(\bm{\Sigma},{m})$ with elements of $\bm{\Sigma}$ as $\sigma_{ij}$.
\end{proof}
\begin{algorithm}[t!]
\small
\caption{MUQ in Koopman Operator via EDMD}\label{Al2}
\begin{algorithmic}
\State \textbullet\ Obtain the data $\mathcal{D}_{obs}=\{\mathbf{X}_{obs},\mathbf{Y}_{obs}\}$
\State \textbullet\ Construct the feature matrix of the inputs $\bm{G}$ and outputs $\bm{A}$ using the basis function $\bm{\psi}(\mathbf{X}_{obs})$ and $\bm{\psi}(\mathbf{Y}_{obs})$
\State \textbullet\ Calculate the variance matrix $\bm{\Sigma}=\textrm{diag}(\sigma_{1}^2,\hdots,\sigma_{n}^2,3\sigma_{1}^{4},\hdots,3\sigma_{n}^{4})$
\State \textbullet\ Generate $N$ random samples around $\psi(X_{obs})_{ij},\Phi(Y_{obs})_{ij}$ following the probability distribution $\psi(X)_{ij}\sim \mathcal{N}(\psi(X_{{obs})_{ij}},\Sigma_{jj}^2)$ and $\psi(Y)_{ij}\sim \mathcal{N}(\psi(Y_{{obs})_{ij}},\Sigma_{jj}^2)$
\State \textbullet\ Initialize $k=1,i=1,j=1$
\For {$i:1:2n$}  
   \For {$j:1:2n$} 
\State {$\boldsymbol{-}$} $S(i,j)=\frac{\Sigma_{jj}^{2}}{\Sigma_{ii}^{2}({m}-2n-1)}$;

\If{$i\neq j $}
    \State {$\boldsymbol{-}$}$ S(i,j)= \frac{1}{(m-2n-1)}$   
\EndIf
\State {$\boldsymbol{-}$} $j = j+1 $
\EndFor
\State {$\boldsymbol{-}$} $ i = i+1 $
\EndFor
\State \textbullet\ $\mathbb{E}[K_{ij}]=K_{{obs}_{ij}}$
\State \textbullet\ $\textrm{Var}[K_{ij}]=S_{ij}$
\end{algorithmic}
\end{algorithm}
\begin{theorem}\label{th2}
     As $n \to \infty, {m} \to \infty$, the probability distribution of $\mathbf{X}^{\dag}$ approaches the Gaussian distribution, that is,  $X_{ij}^{\dag}\sim \mathcal{N}(0,\frac{1}{\sigma_{ij}^{2}({m}-n-1)})$.
\end{theorem}
\begin{figure}[t!]
    \centering \includegraphics[height=6cm,width=9cm]{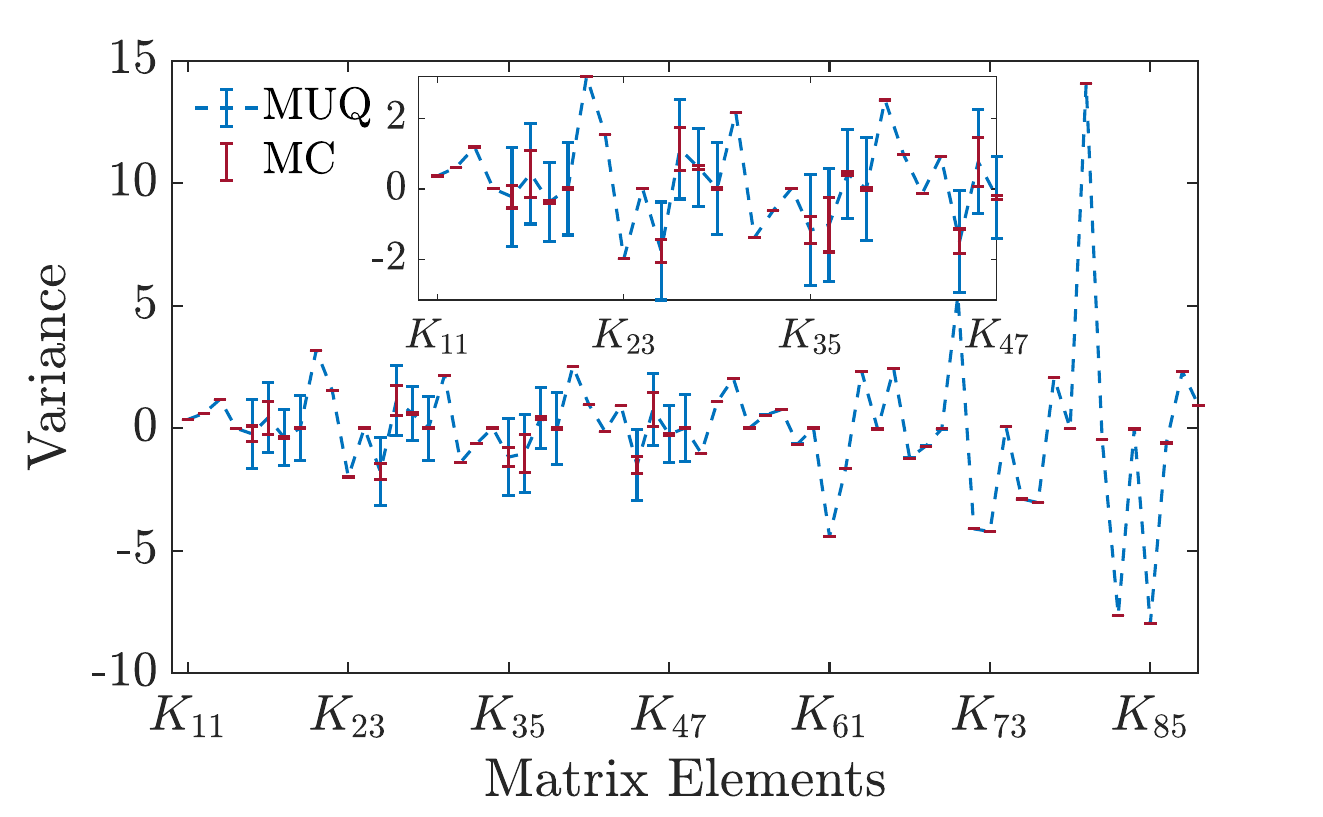}
    \caption{Comparison of results obtained from the uncertainty quantification via the EDMD algorithm for a single-machine infinite-bus system.}
    \label{1edmd}
\end{figure}

We standardize the data matrix, $\mathcal{D}_{obs}=\{\mathbf{X}_{obs},\mathbf{Y}_{obs}\}$, around zero mean to be able to apply Theorem \ref{th1} and Theorem \ref{th2} \cite{Cook2011OnMatrix}. Let us define the standardized random variables as $\Tilde{\bm{X}}=\bm{X}-\mathbf{X}_{obs}$ and $\Tilde{\bm{Y}}=\bm{Y}-\mathbf{Y}_{obs}$, respectively. Consequently, their elemental distribution is given by $\Tilde{X}_{ij}\sim\mathcal{N}(0,\sigma_{j}^{2})$ and $\Tilde{Y}_{ij}\sim\mathcal{N}(0,\sigma_{j}^{2})$. Let us construct the feature matrix, $\Tilde{\bm{G}}=[\bm{\psi}(\Tilde{x}_{1}),\hdots,\bm{\psi}(\Tilde{x}_{m})].$ Similarly, the feature matrix of the outputs can be constructed as $\Tilde{\bm{A}}=[\bm{\psi}(\Tilde{y}_{1}),\hdots,\bm{\psi}(\Tilde{y}_{m})]$. 
Let us denote: 

\begin{equation*}
\Tilde{\mathbf{G}}^{\dag}=\begin{bmatrix}
\Tilde{G}_{11}^{\dag} & \hdots & \Tilde{G}_{1{m}}^{\dag}\\
\vdots &   \ddots  &  \vdots\\
\Tilde{G}_{2n,1}^{\dag} & \hdots  & \Tilde{G}_{2n,{m}}^{\dag}\\
\end{bmatrix}.    
\end{equation*}
\begin{figure}[t!]%
\centering
\subfloat[\centering ]{{\includegraphics[height=2cm,width=2.9cm]{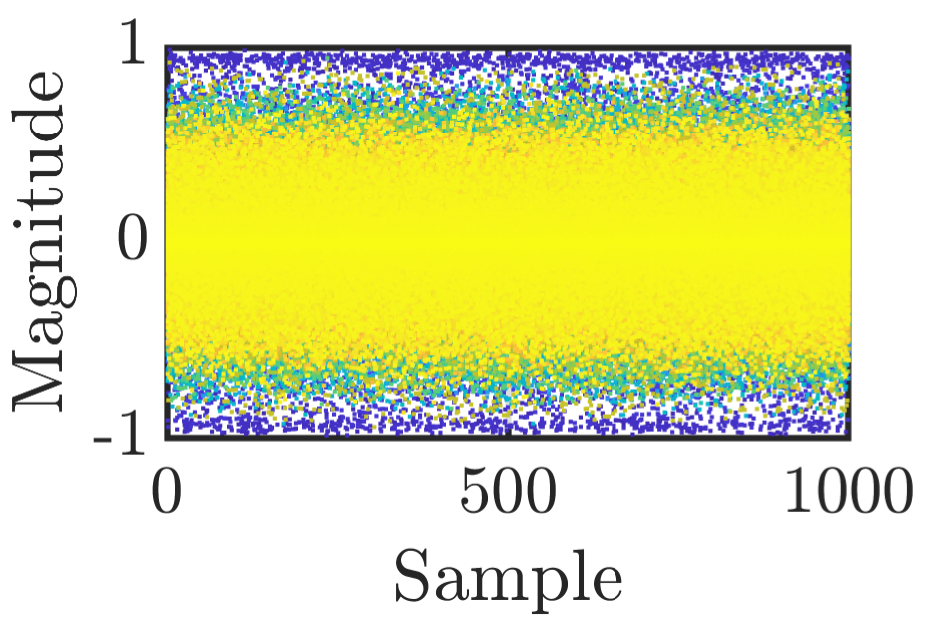} }}%
\subfloat[\centering ]{{\includegraphics[height=2cm,width=2.9cm]{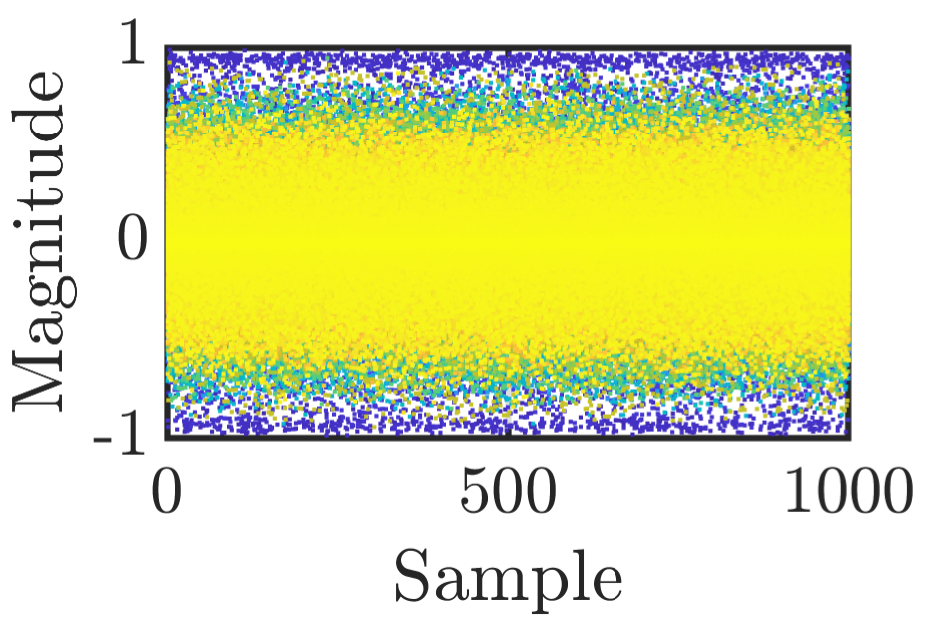} }}%
 \subfloat[\centering ]{{\includegraphics[height=2cm,width=2.9cm]{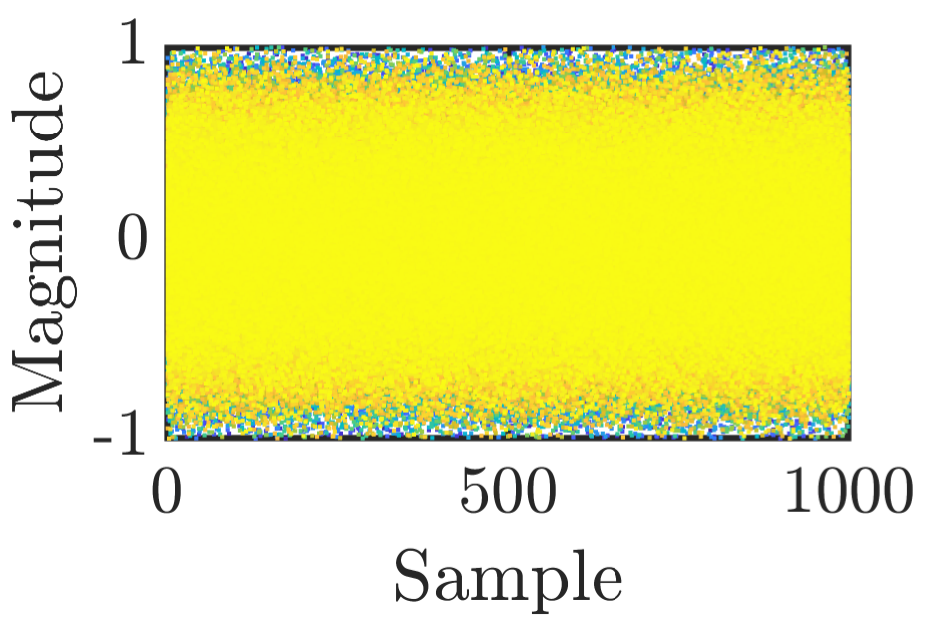}}}%
\caption{{Comparison between the true samples obtained from Monte Carlo for the (a) Koopman modes and (b) eigenfunctions and (c) the ones obtained from the proposed MUQ of the eigenfunctions for the 3-Bus system. Distinct hues represent the values across various samples.}}%
\label{4_eigvec}
\end{figure}%
From Theorem \ref{th1}, we get the first two moments of the pseudo-inverse of G as:
\begin{equation}
    \mathbb{E}[\Tilde{G}^{\dag}_{ij}]=0,
\end{equation}
and: 
\begin{equation}
\mathbb{E}[(\Tilde{G}^{\dag}_{ij})^{2}]=\frac{1}{\sigma_{i}^{2}m({m}-2n-1)},
\end{equation}
respectively.
Delving into the EDMD algorithm for the Koopman operator estimation, we
define the new random variables for each element of the Koopman operator matrix as: 
\begin{equation}
{\Tilde{\mathbf{K}} = \begin{bmatrix}
             \Tilde{K}_{11} &    \hdots & \Tilde{K}_{1,2n}\\
            \vdots  &   \ddots  & \vdots \\
             \Tilde{K}_{2n,1} &  \hdots &  \Tilde{K}_{2n,2n} 
         \end{bmatrix},\\}
\end{equation}
\begin{equation*}
        \resizebox{1\hsize}{!}{$= \begin{bmatrix}
\Tilde{G}_{11}^{\dag}\Tilde{A}_{11}+\hdots+\Tilde{G}_{1{m}}^{\dag}\Tilde{A}_{{m}1} & \hdots & \Tilde{G}_{11}^{\dag}\Tilde{A}_{1,2n}+\hdots+\Tilde{G}_{1{m}}^{\dag}\Tilde{A}_{{m},2n}\\ 
             \vdots    & \ddots &\vdots\\
        \Tilde{G}_{2n,1}^{\dag}\Tilde{A}_{11} +\hdots  +\Tilde{G}_{2n,{m}}^{\dag}\Tilde{A}_{{m}1} & \hdots & \Tilde{G}_{2n,1}^{\dag}\Tilde{A}_{1,2n} +\hdots +\Tilde{G}_{2n,{m}}^{\dag}\Tilde{A}_{{m},2n} 
    \end{bmatrix}$}.
\end{equation*}
Let us consider the diagonal elements of the Koopman operator, $\Tilde{\mathbf{K}}$. The entries of $\Tilde{K}_{ii}$ involve the product of the two independent random variables, $\Tilde{G}^{\dag}_{it}\sim \mathcal{N}\left(0,\frac{1}{\Sigma_{ii}^{2}({T}-2n-1)}\right)$ and $\Tilde{A}_{ti}\sim \mathcal{N}\left(0,{\Sigma_{ii}^{2}}\right)$. Their first and second moments can be derived as:  
\begin{align}
\mathbb{E}[\Tilde{K}_{ii}] &=  \mathbb{E}[\Tilde{G}_{i1}^{\dag}\Tilde{A}_{1i}+\hdots+\Tilde{G}_{im}^{\dag}\Tilde{A}_{Ti}],\\
&= \mathbb{E}[\Tilde{G}_{11}\Tilde{A}_{11}] + \hdots+ \mathbb{E}[\Tilde{G}_{1m}\Tilde{A}_{m1}],\\
&=  \mathbb{E}[\Tilde{G}_{i1}]  \mathbb{E}[\Tilde{A}_{1i}] +\hdots+ \mathbb{E}[\Tilde{G}_{im}] * \mathbb{E}[\Tilde{A}_{mi}],\\
&=0,
\end{align}
\begin{align}
\mathbb{E}[\Tilde{K}_{ii}^{2}] &= \mathbb{E}[(\Tilde{G}_{i1} \Tilde{A}_{1i} + \hdots+ \Tilde{G}_{im}  \Tilde{A}_{mi})^{2}],\\ 
 &= \mathbb{E}[(\Tilde{G}_{i1}  \Tilde{A}_{1i})^{2}] + \hdots+  \mathbb{E}[(\Tilde{G}_{im}  \Tilde{A}_{Ti})^{2}],\\
&= \mathbb{E}[(\Tilde{G}^{\dag}_{i1})^{2}]  \mathbb{E}[\Tilde{A}_{1i}^2] + \hdots+ \mathbb{E}[(\Tilde{G}^{\dag}_{im})^{2}]  \mathbb{E}[\Tilde{A}_{mi}^2],\\
&= \frac{1}{({m}-2n-1)},
\end{align}
respectively.

Consider an off-diagonal element, $K_{ij}$, which involves the sum of the product of random variables $\Tilde{G}^{\dag}_{it}\sim \mathcal{N}(0,\Sigma_{ii}^{2})$ and $\Tilde{A}_{tj}\sim \mathcal{N}(0,\frac{1}{\Sigma_{jj}^{2}})$. Its first and second moment can be developed as:
\begin{align}
\mathbb{E}[\Tilde{K}_{ij}] &= \mathbb{E}[\Tilde{G}_{i1}^{\dag} \Tilde{A}_{1j}+\hdots+\Tilde{G}_{im}^{\dag}\Tilde{A}_{mj}],\\
&= \mathbb{E}[\Tilde{G}_{i1}^{\dag}\Tilde{A}_{1j}] + \hdots+ \mathbb{E}[\Tilde{G}_{im}^{\dag}\Tilde{A}_{mj}],\\
&=  \mathbb{E}[\Tilde{G}_{i1}^{\dag}]  \mathbb{E}[\Tilde{A}_{1j}] +\hdots+ \mathbb{E}[\Tilde{G}_{im}^{\dag}] *\mathbb{E}[\Tilde{A}_{mj}],\\
    &=0,
\end{align}
\begin{align}
\mathbb{E}[\Tilde{K}_{ij}^{2}] &= \mathbb{E}[(\Tilde{G}_{i1} \Tilde{A}_{1j} + \hdots+ \Tilde{G}_{im}  \Tilde{A}_{mj})^{2}],\\ 
&= \mathbb{E}[(\Tilde{G}_{i1}^{\dag}  \Tilde{A}_{1j})^{2}] + \hdots+  \mathbb{E}[(\Tilde{G}_{im}^{\dag}  \Tilde{A}_{mj})^{2}],\\
&= \mathbb{E}[(\Tilde{G}^{\dag}_{i1})^{2}] \mathbb{E}[\Tilde{A}_{1j}^2] + \hdots+ \mathbb{E}[(\Tilde{G}^{\dag}_{im})^{2}]  \mathbb{E}[\Tilde{A}_{mj}^2],\\
&= \frac{\Sigma_{jj}^{2}}{\Sigma_{ii}^{2}({m}-2n-1)},
\end{align}
respectively.
\begin{figure}[t!]%
\centering
\subfloat[\centering ]{{\includegraphics[height=4cm,width=4.5cm]{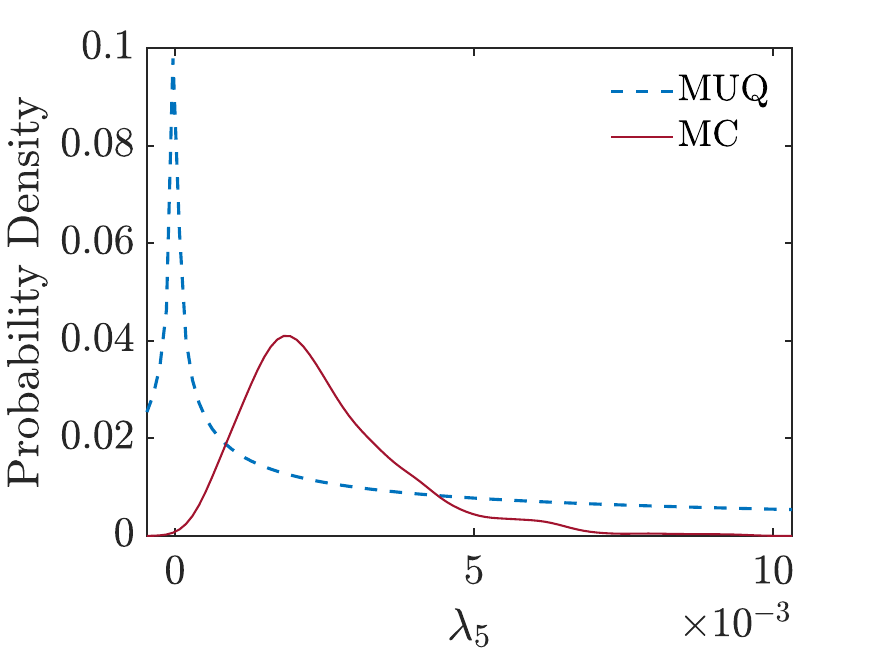} }}%
\subfloat[\centering ]{{\includegraphics[height=4cm,width=4.5cm]{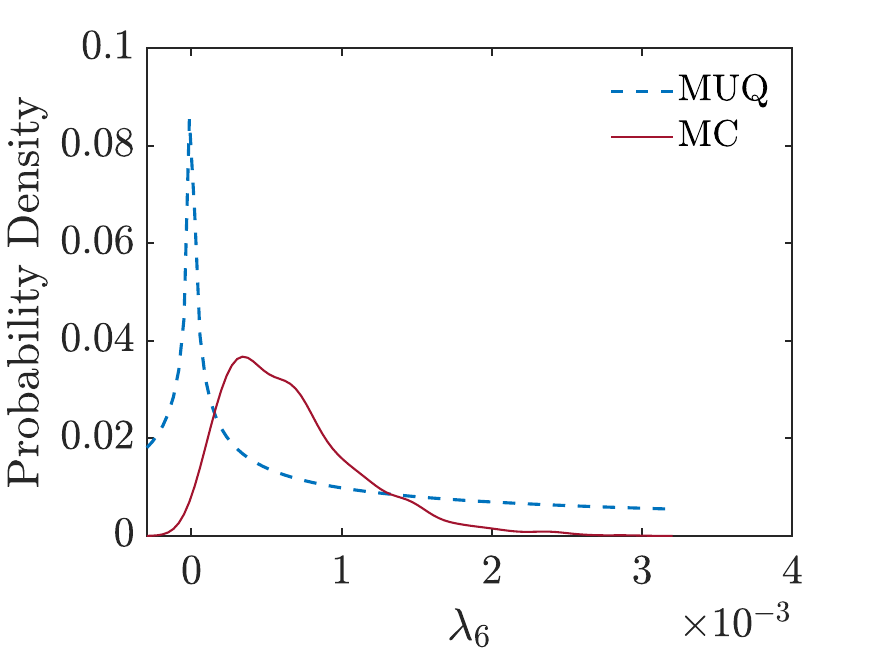} }}%
\quad
\subfloat[\centering ]{{\includegraphics[height=4cm,width=4.5cm]{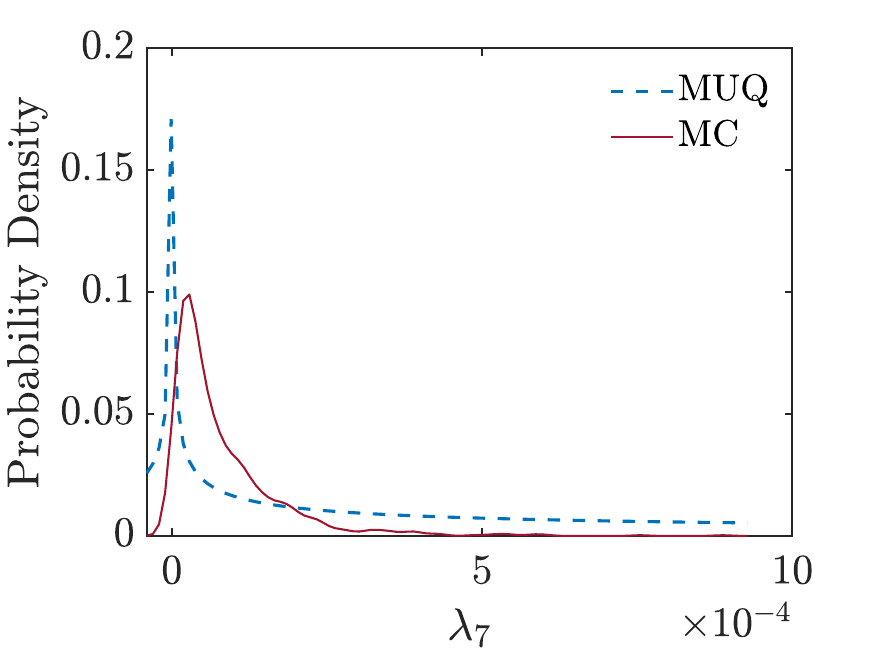} }}%
\subfloat[\centering ]{{\includegraphics[height=4cm,width=4.5cm]{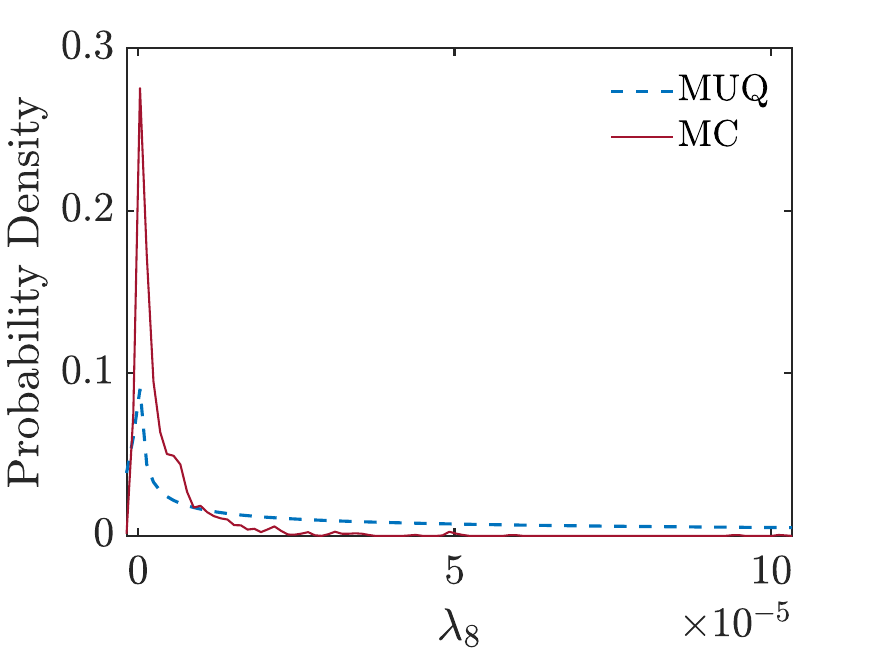} }}%
\caption{Comparison between the probability densities obtained from Monte Carlo and the ones obtained from the proposed uncertainty quantification MUQ of the Koopman eigenvalues (a) $\lambda_{5}$, (b) $\lambda_{6}$, (c) $\lambda_{7}$, and (d) $\lambda_{8}$ for the 3-Bus system.}%
    \label{4_eigval_sm}
\end{figure}%
\begin{figure}[t!]%
\centering
\subfloat[\centering ]{{\includegraphics[height=4cm,width=4.5cm]{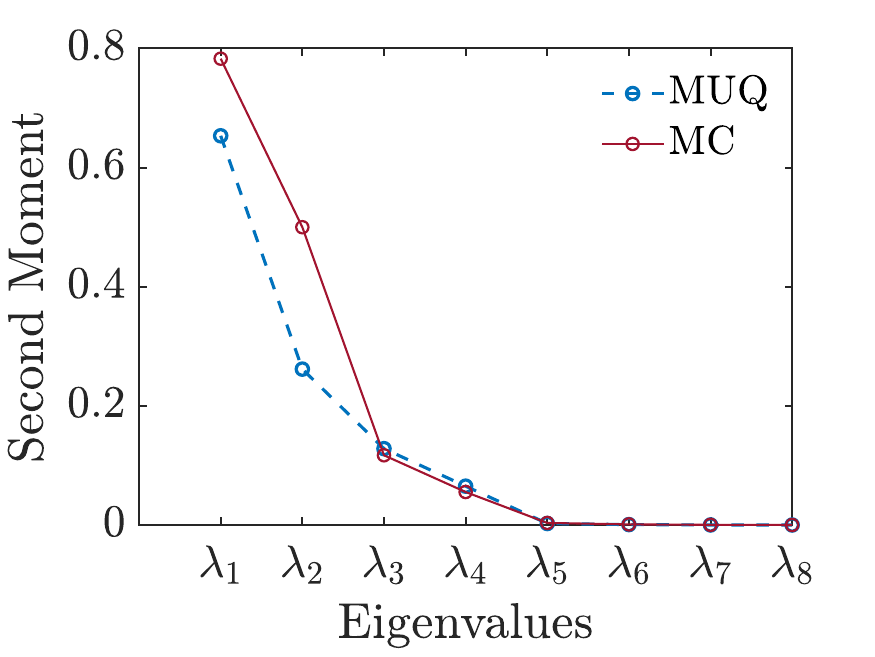} }}%
\subfloat[\centering ]{{\includegraphics[height=4cm,width=4.5cm]{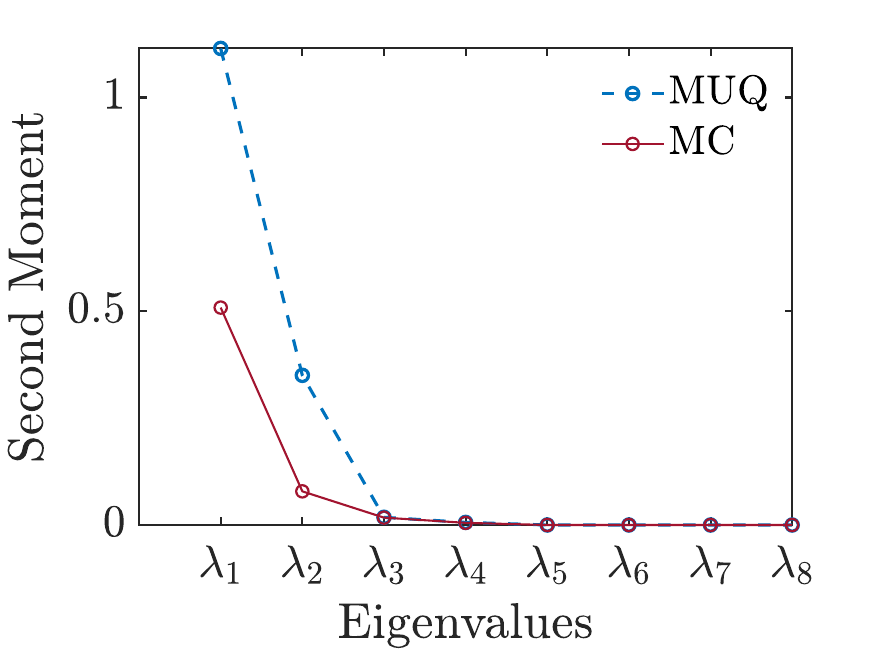}}}%
\caption{{Comparison of the (a) first and (b) second moments of the $n=8$ Koopman eigenvalues of the Koopman operator obtained from Monte Carlo and the proposed MUQ for the 3-Bus system.}}%
 \label{moments_sm}
\end{figure}%
Let us gather these findings to obtain the first and second moments of each element of $\mathbf{K}$. The first moment defined by the expectation of the $ij^{th}$ element of the random matrix, $\mathbf{K}$, is given by: 
\begin{equation}
        \mathbb{E}[{\Tilde{K}_{ij}}]=0.
\end{equation}
\begin{figure}[ht!]%
\centering
{\includegraphics[height=6.5625cm,width=8.9370cm]{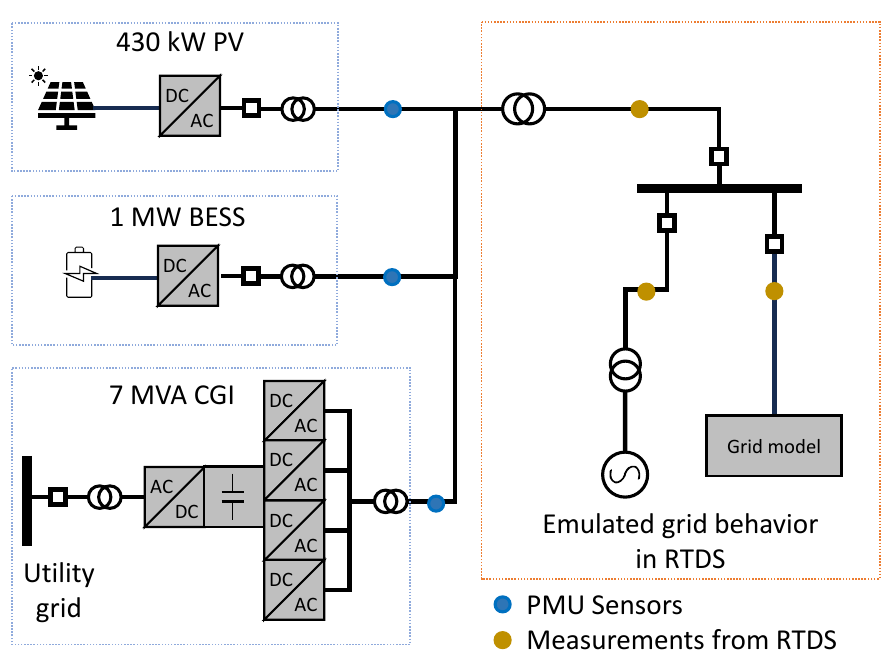}}%
\caption{Power-Hardware-In-Loop (PHIL) setup for grid emulation}%
    \label{fig_ARIES_grid}%
\end{figure}
\begin{figure}[h!]%
\centering
\subfloat[\centering]{{\includegraphics[height=3cm,width=4.7cm]{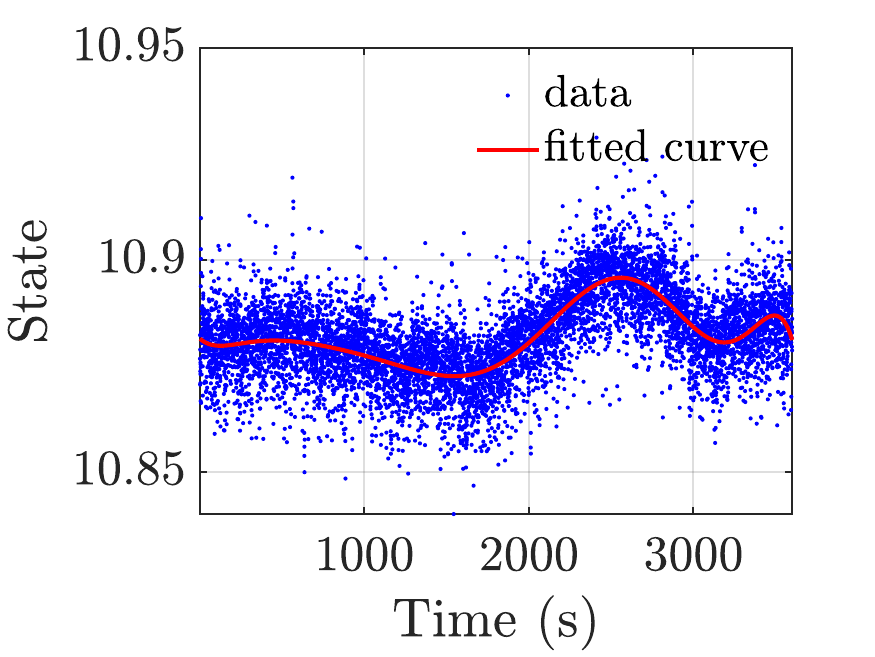}}}%
\subfloat[\centering]{{\includegraphics[height=3cm,width=4.6cm]{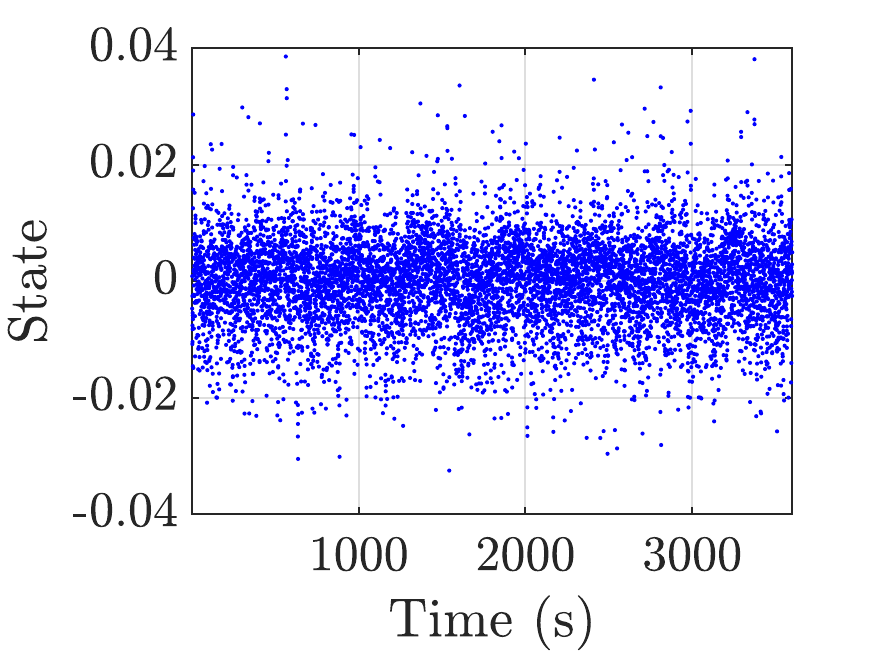}}}%
\caption{{(a) The data plot for State 1 and (b) the data with subtracted regressed values.}}%
    \label{data_rw1}%
\end{figure}

The second moment defined by the variance of the element $ij^{th}$ of the random matrix, $\mathbf{K}$, is expressed as:
\begin{equation}\label{var}
\mathbb{E}[\Tilde{K}_{ij}^{2}] =
\begin{cases} 
\frac{\Sigma_{jj}^{2}}{\Sigma_{ii}^{2}({m}-2n-1)}\; i\neq j,\\
 \frac{1}{({m}-2n-1)}\; i= j.
\end{cases} 
\end{equation}

\begin{figure*}[ht!]%
\centering
\subfloat[\centering]{{\includegraphics[height=4.6cm,width=4.5cm]{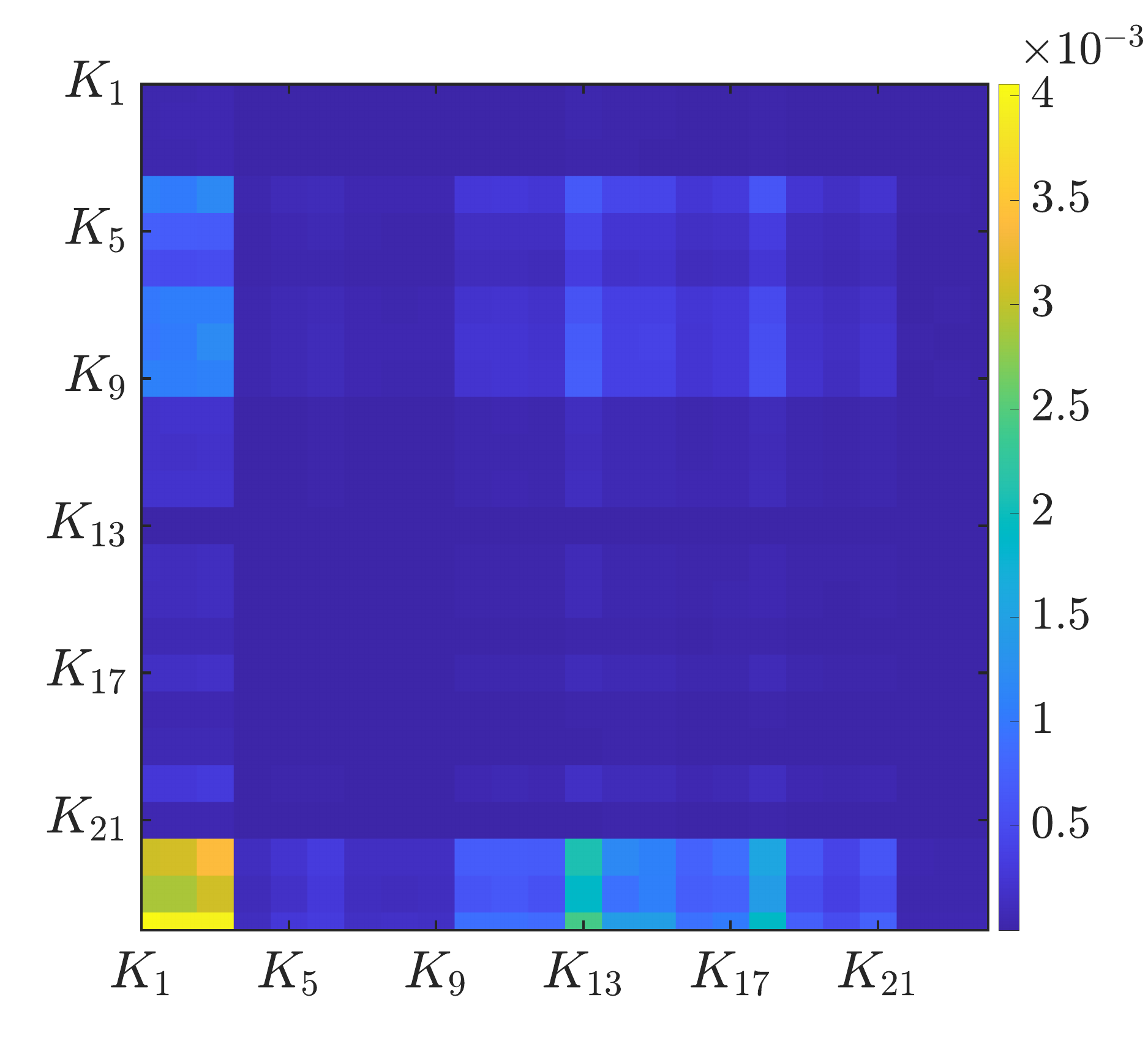} }}%
\subfloat[\centering]{{\includegraphics[height=4.6cm,width=4.5cm]{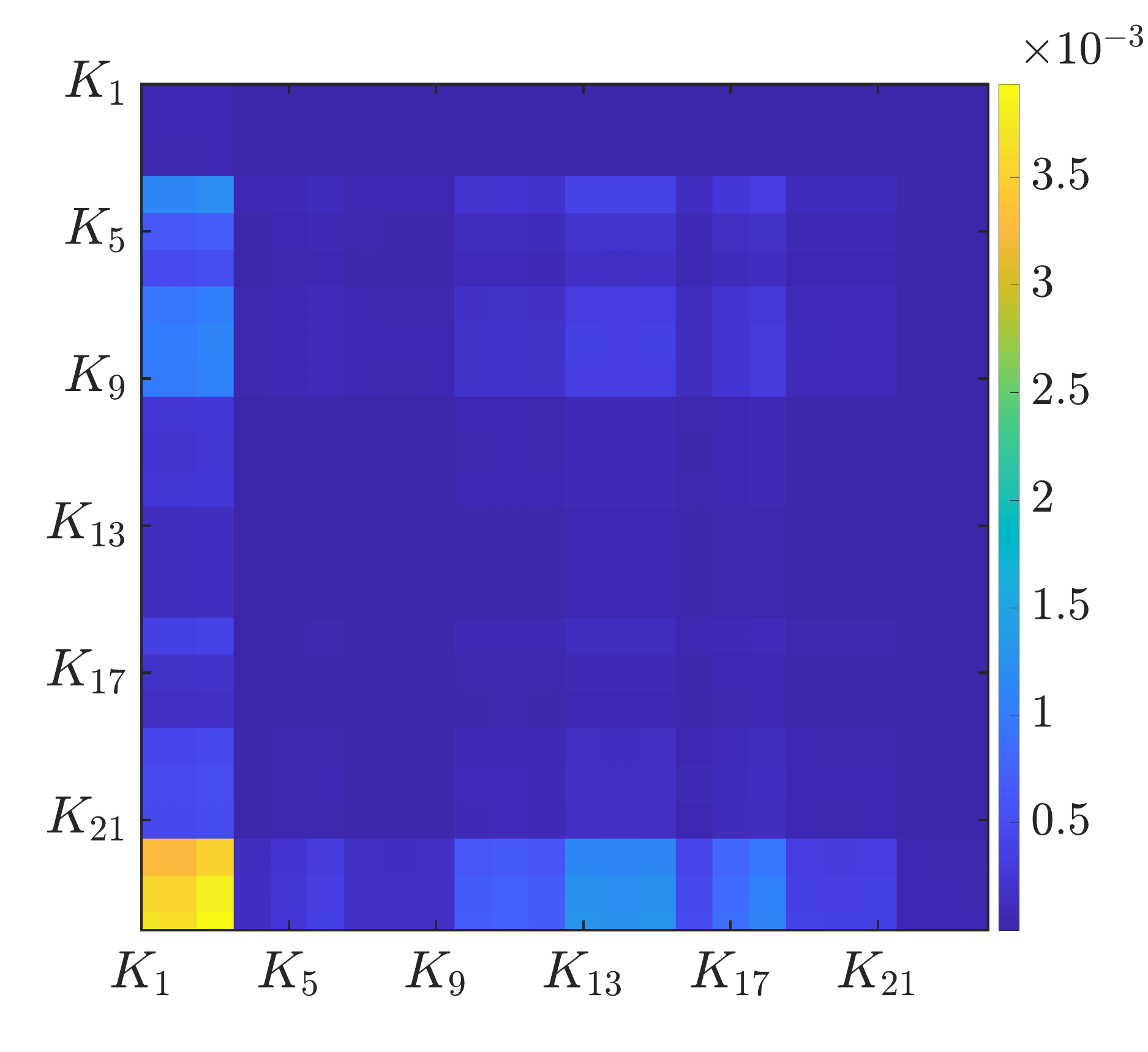} }}%
 \subfloat[\centering]{{\includegraphics[height=4.6cm,width=4.5cm]{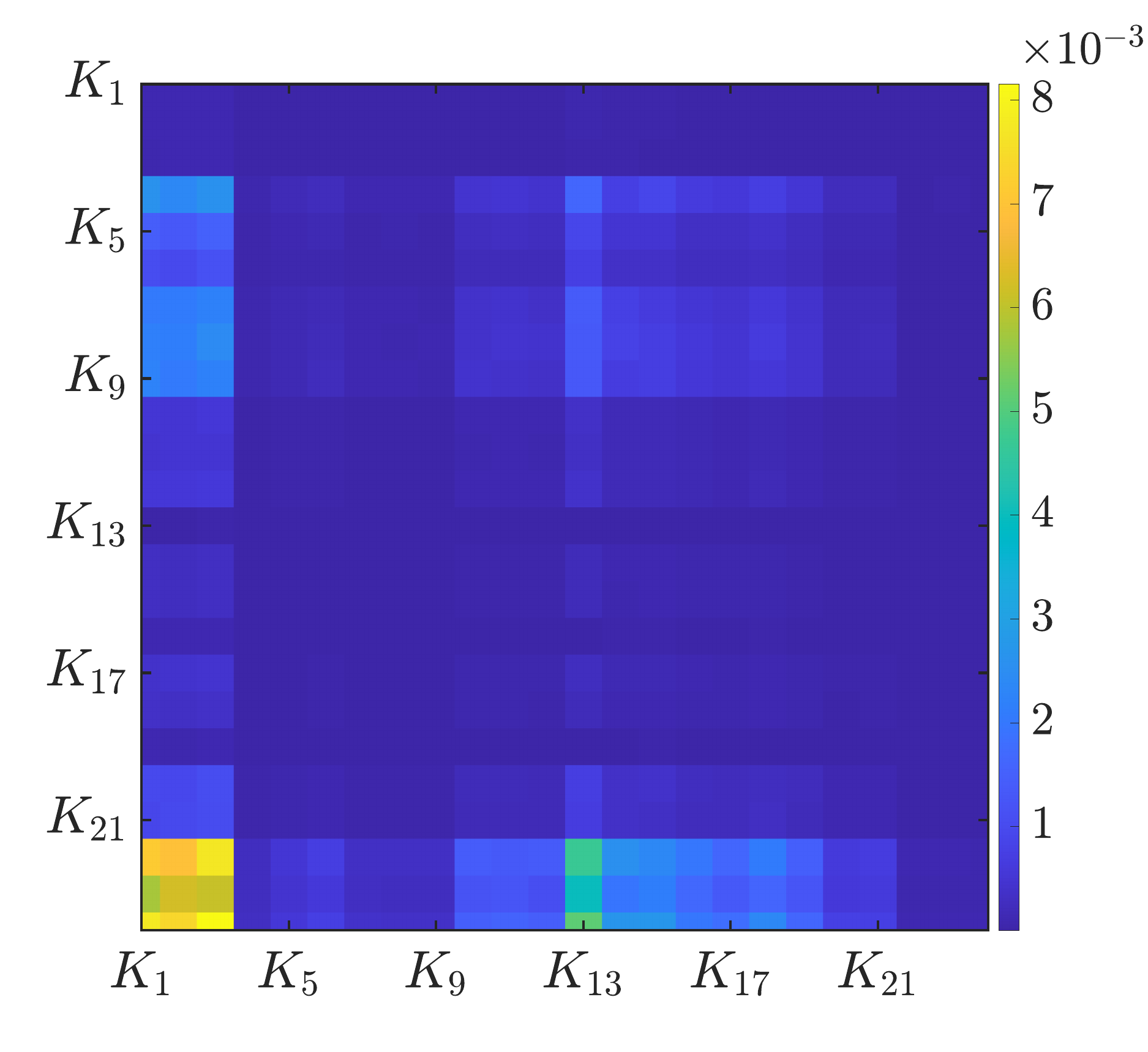}}}%
\subfloat[\centering]{{\includegraphics[height=4.6cm,width=4.5cm]{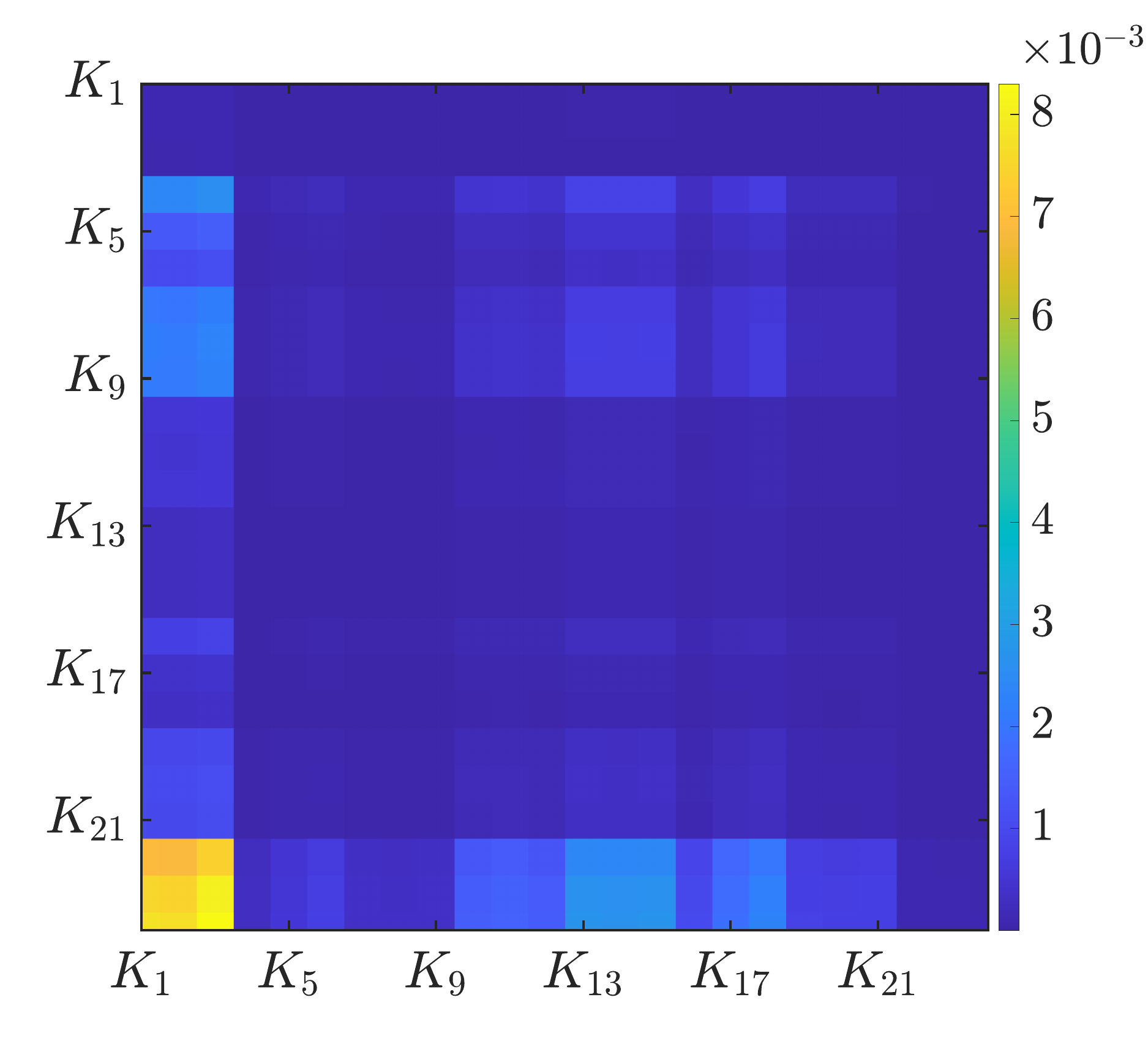} }}%
\caption{{Comparison between (a) the true variance, $R$, obtained from Monte Carlo and (b) the ones obtained from the proposed MUQ algorithm, $S$, via DMD; (c) variances obtained from Monte Carlo and (d) MUQ via EDMD.}}%
    \label{rw}%
\end{figure*}

We are interested in the expectation of the Koopman operator for the data $\mathcal{D}=\{\mathbf{X},\mathbf{Y}\}$. To transform our analysis back from $\Tilde{\mathcal{D}}$ to ${\mathcal{D}}$, we employ Theorem \ref{th2}, which states that the newly defined random variables of the Koopman operator approach the Gaussian distribution as $n\to \infty; T\to \infty.$ This is compatible with estimation algorithms, such as EDMD, in which the complex dynamics with a Koopman operator are accommodated by enhancing the observable function space with location-equivariant and location-invariant statistics.  We know that the mean is a location-equivariant statistic \cite{Lopuhaa1991BreakdownMatrices}. The mean of the elements of the Koopman operator that are used for the data, $\mathcal{D}$, is given by:
\begin{equation}
    \mathbb{E}[K_{ij}]=K_{{obs}_{ij}},
\end{equation}
with the variance $\mathbb{E}[K_{ij}^{2}]=\mathbb{E}[\Tilde{K}_{ij}^{2}]$ given by \eqref{var}. The steps for uncertainty quantification via the estimation of the EDMD Koopman operator are described in Algorithm \ref{Al2}. The proposed MUQ algorithm used for the estimation of the DMD Koopman operator is described in Algorithm \ref{Al1}.
\subsection{Probability Distribution of Koopman Tuples}
The Koopman modes, eigenfunctions, and Koopman eigenvalues are distributed independently of each other. As a result, the correlation between the matrix elements and the Koopman eigenvalues vanishes. Let us prove this by considering $\mathbb{E}[K_{ij}v_{j}]$, where $v_{j}$ is the $j^{th}$ eigenfunction of the Koopman operator. We express the $ij^{th}$ element of the Koopman operator in terms of its eigenfunction and Koopman eigenvalues as:
Let us consider the product of $K_{ij}$ with $v_{j}$, which can also be given as: 
\begin{equation}
        K_{ij}v_{j}=\sum_{k} V_{ki}\lambda_{k}V_{kj}v_{1j},
\end{equation}
The expectation of which vanishes as $\mathbb{E}[\lambda_{k}]=0.$ Therefore:
\begin{equation}
    \mathbb{E}[K_{ij}v_{j}]=0.
\end{equation}
We will use this property to draw conclusions about the probability distributions of the Koopman modes and eigenfunctions of the Koopman operator.

\subsection{Probability Distribution of the Eigenvalues of the Koopman Operator}
The eigenvalues of a symmetric random matrix follow the Marchenko-Pastur distribution \cite{Marcenko1967DistributionMatrices}, expressed as:
\begin{equation}\label{mp}
f(x;\lambda, \sigma^2) = \frac{\sqrt{(\lambda_+ - x)(x - \lambda_-)}}{2 \pi \sigma^2 x},
\end{equation}
with $\lambda_{\pm}=\sigma^{2}(1\pm\sqrt{\lambda})^{2}.$
Here, $x$ is the variable, $\lambda_+$ and $\lambda_-$ are the upper and lower bounds of the spectrum, and $\sigma^2$ is the variance of the entries of the random matrix. First, we will use this knowledge to construct the eigenvalues of the symmetric matrix, constructed as $\bm{K}_{sym}=\frac{\bm{K}\bm{K}'}{n}$. The probability distribution of the eigenvalues of the random matrix, $\bm{K}_{sym}$, can easily be shown to be given by the probability distribution defined in \eqref{mp}. 

To derive the distribution of the Koopman eigenvalues $\lambda_{1},\hdots,\lambda_{n}$ of $\bm{K}$, which is the quantity of our interest, we take a positive square root of the eigenvalues $s_{1},\hdots,s_{n}$ of $\bm{K}_{sym}$.
Note that this article specifically deals with the real parts of the Koopman eigenvalues using the Marchenko-Pastur distribution for the squared random eigenvalues of the Koopman operator, $\bm{K}$.

\subsection{Probability Distribution of the Koopman Eigenfunctions and Koopman Modes}
Given that its correlations with the matrix elements are zero, the probability distribution of the Koopman eigenfunctions and Koopman modes scales only with $n$. As the matrix dimensions, $n$, tend to infinity, the Koopman eigenfunctions and modes scale as $\frac{1}{n}$ in a limiting distribution \cite{Bai2007OnMatrix}.
For finite matrix dimensions, the Koopman eigenfunctions and modes of the random matrix, $\bm{K}$, are uniformly distributed with the Haar measure, $O(n)$ \cite{Silverstein1981DescribingGroups}. The Haar measure ensures that each matrix dimension is orthogonal to all previous dimensions. As we sample with the Haar measure, each new eigenvector and its random realizations must be orthogonal to all previous directions.

Computationally, this can be achieved using the QR decomposition of the set of random samples following a normal distribution with zero mean and unit variance. The QR decomposition breaks down a matrix, $A$, into a product, $A = QR$, of an orthonormal matrix, $Q$, and an upper triangular matrix, $R$.
Note that this holds true for both normalized Koopman modes and eigenfunctions of the Koopman operator.

\begin{figure*}[t!]%
\centering
\subfloat[\centering ]{{\includegraphics[height=5cm,width=4.65cm]{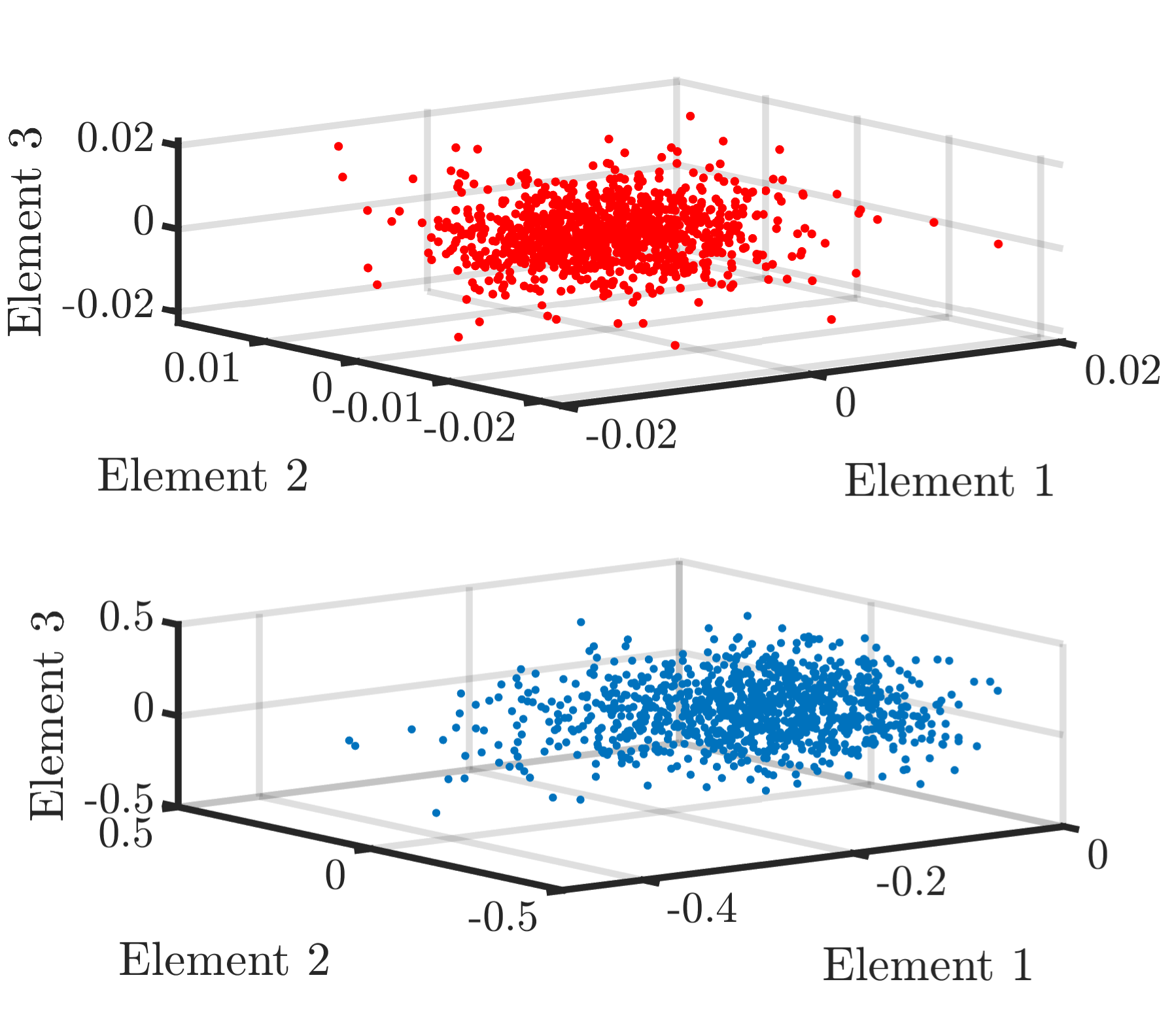}}}%
\subfloat[\centering ]{{\includegraphics[height=5cm,width=4.65cm]{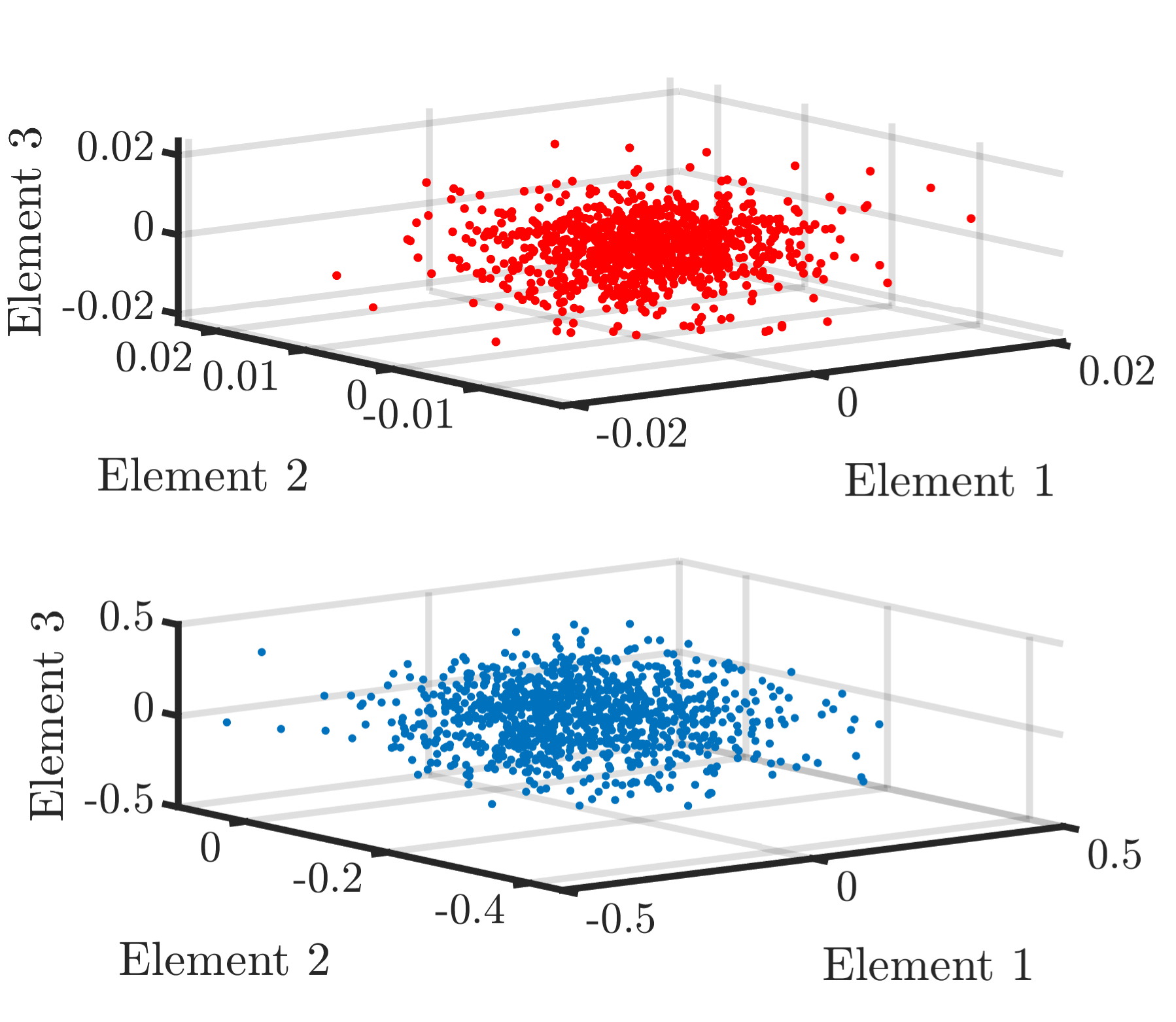}}}%
 \subfloat[\centering ]{{\includegraphics[height=5cm,width=4.65cm]{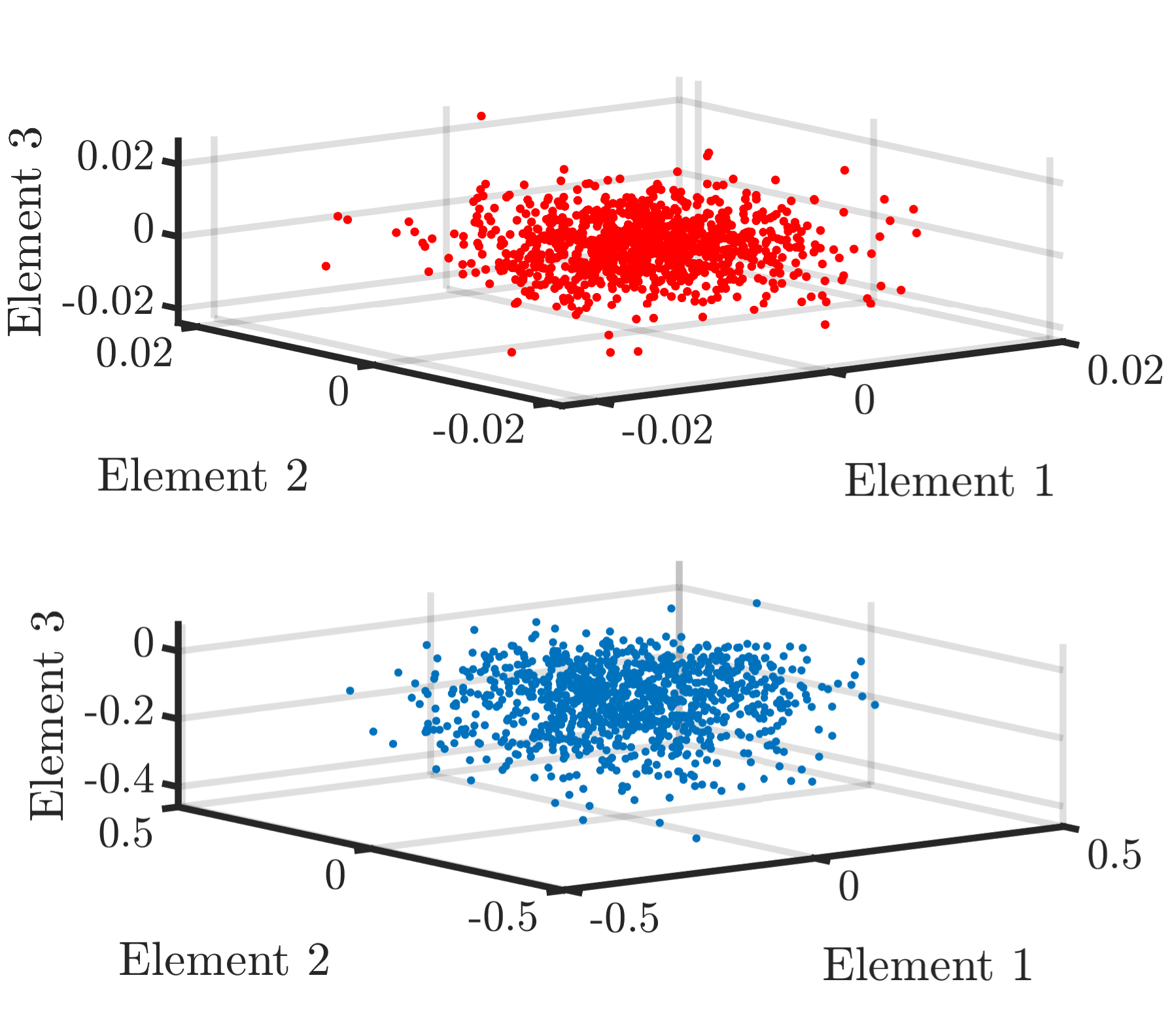}}}%
  \subfloat[\centering]{{\includegraphics[height=5cm,width=4.65cm]{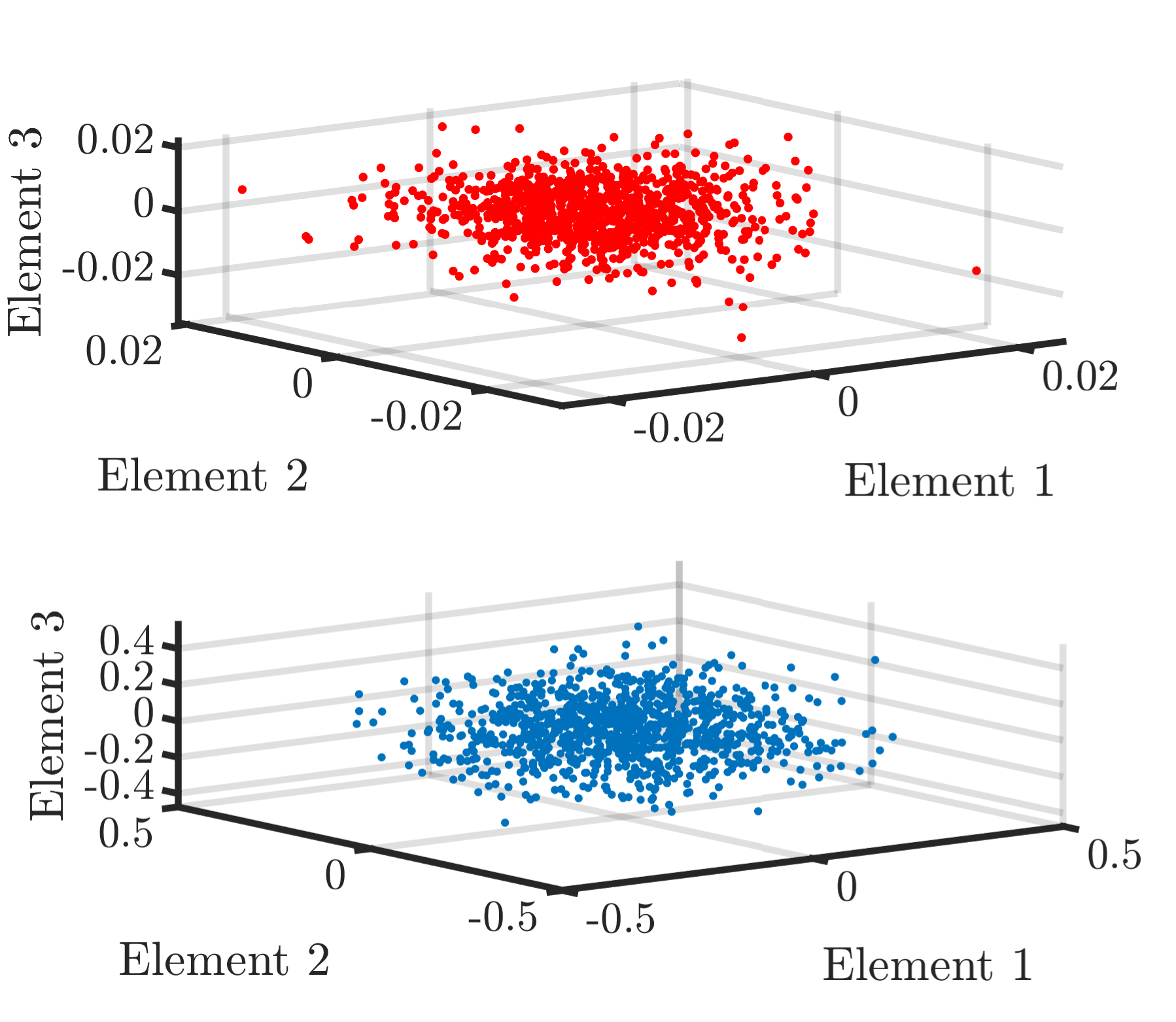}}}%
\caption{Comparison of the first three elements of the (a) first, (b) second, (c) third, and (d) fourth eigenfunction between the true samples obtained from the Monte Carlo simulation and from the proposed uncertainty quantification method of the eigenfunctions indicated in red and blue colors respectively.}%
    \label{dimensions}
\end{figure*}%
\section{Numerical Analysis}\label{sec4}
In this section, we discuss the implementation of the proposed uncertainty quantification in the data-driven identification of nonlinear dynamics using measurements from both a simulated example power system and real-world data.
\subsection{Three-Bus Example} 
Consider a power network with two generators and a load, as shown in Figure \ref{3_bus_fig} (a).  
The load is modeled as an induction motor in parallel with a constant $PQ$ load. The system is modeled as a four-dimensional dynamical system with the states being generator angle $(\delta_2)$, generator angular velocity $(\omega)$, load angle $(\delta_3)$, and load voltage magnitude $(V_3)$. For a detailed analysis of the dynamics and parameters of the system, we refer the reader to \cite{dobson_model, ajjarapu_bifurcation}.

Let us obtain the measurements of the $n=4$ dynamic states $\{\delta_2, \omega, \delta_3, V_3 \}$ by solving the continuous differential equations of the given system for the duration ${T}$. We stack these so-called observed quantities in the data matrix, $\mathcal{D}_{obs}\in \mathbb{R}^{{T}\times n}= 
[\bm{\delta}_2, \bm{\omega}, \bm{\delta}_3, \bm{V}_3]'$, shown in Figure \ref{3_bus_fig} (b). Let us form the observed input matrix using the $T-1$ rows of the matrix $\mathbf{D}$ as $\mathbf{X}_{obs}\in\mathbb{R}^{{T}-1\times n}=[\mathbf{d}_{1},\hdots,\mathbf{d}_{{T}-1}]$ and the observed output matrix $\mathbf{Y}_{obs}\in\mathbb{R}^{{{T}}-1\times n}=[\mathbf{d}_{2},\hdots,\mathbf{d}_{{T}}]$. 
The Koopman operator is estimated using the observations $\mathcal{D}_{obs}=\{\mathbf{X}_{obs},\mathbf{Y}_{obs}\}$ from the DMD algorithm as $\mathbf{K}_{obs}=\mathbf{X}_{obs}^{\dag}\mathbf{Y}_{obs}$. 

Let us now characterize the measurement uncertainty using the variance matrix, $\bm{\Sigma}$, calculated from the steady-state measurements from the time, approximately $5$ s to $10$ s, which is given as 
    $\bm{\Sigma}=\textrm{diag}(0.39745,0.0026435,0.5936,0.23668).$
Now, with the variance, $\bm{\Sigma}$, and mean, $\mathbf{X}_{obs}$, i.e., ${X}_{ij}\sim\mathcal{N}(X_{{obs}_{ij}},\Sigma_{jj})$ and ${Y}_{ij}\sim\mathcal{N}(Y_{{obs}_{ij}},\Sigma_{jj})$, we can draw $N=1000$ random samples, $\mathcal{D}^{(k)}=\{\bm{X}^{(k)},\bm{Y}^{(k)}\}_{k=1}^{N}$, to conduct a Monte Carlo simulation. 
The realizations of the Koopman operator, $\bm{K}^{(k)}$; $k=1,\hdots,N$, are obtained by employing the DMD algorithm as follows: 
\begin{equation}
    \bm{K}^{(k)}= \bm{X}^{{\dag}^{(k)}}\bm{Y}^{(k)}.
\end{equation}
The variance of the elements of the Koopman operator, $\hat{R}_{ij}=\textrm{Var}(K_{ij})=\sum_{k=1}^{N}\frac{({K}_{ij}^{{(k)}}-K_{{obs}_{ij}})^{2}}{N-1}$, is estimated using the sample variance, which represents the true variance of the Koopman operator for comparison. We employ Algorithm \ref{Al1} to estimate the expectation $\mathbf{K}_{obs}$ and variances of the elements of the Koopman operator, $\mathbf{S}$. The Koopman operator is sampled from the distribution $K_{ij}\sim \mathcal{N}(K_{obs},\hat{S}_{ij});\;i=1,\hdots,4;\;j=1,\hdots,4.$
The results are compared with the Monte Carlo method where the elements of the Koopman operator are sampled from $K_{ij}\sim\mathcal{N}(K_{{obs}_{ij}},\hat{R}_{ij}).$ The comparison of the results is shown in Figure \ref{prob_1}. 
Similarly, we also analyze the given system in the function space using polynomial dictionary functions. The results obtained for the EDMD estimation using Algorithm \ref{Al2} are displayed in Figure \ref{1edmd}. As shown, the analytical variance obtained from the proposed MUQ method matches the system realization through the Monte Carlo method. 

Figure \ref{prob_1} 
show that the uncertainty quantification obtained by emplying Algorithm \ref{Al1} is quite close to the variance values obtained from Monte Carlo simulations for the single-machine system.
Figure \ref{1edmd} shows the distributions of the elements of the Koopman operator for the case of EDMD estimation algorithm. Note that the states of the system are doubled in this case. As a result, the sample variance effectively quantifies the elemental uncertainty in the Koopman operator.
\subsection{Results on the Distributions of the Spectral Properties of the Koopman Operator}
For this $4-$D state system, the application of EDMD with quadratic polynomial lifting leads to the dimensions of the Koopman operator, $n=8$. The spherical orthogonal space of the eigenfunctions of the Koopman operator in 2-D is visualized in Figure \ref{4_eigvec}. Note that this plot has all the random eigenfunctions of the $N=1000$ realizations of the Koopman operator. 

As for the Koopman eigenvalues, we compute the first two moments of the $n=8$ eigenvalues from the Monte Carlo method, which are obtained by computing the eigenvalues of the $N=1000$ realizations of the Koopman operator with mean $\mathbf{K}_{obs}$ and the variance matrix, $\mathbf{V}$, computed by employing the proposed MUQ. These Monte Carlo computations of the first two moments of the Koopman eigenvalues are compared with those obtained by employing the Marchenko-Pastur density function as a probability density measure in Figure \ref{moments_sm}. Note that the density values extracted from both Monte Carlo and MUQ are normalized to 1.  
The probability densities of the Koopman eigenvalues are individually compared in Figure \ref{4_eigval_sm}.
\begin{figure}[t!]%
    \centering
\subfloat[\centering ]{{\includegraphics[height=4cm,width=4.5cm]{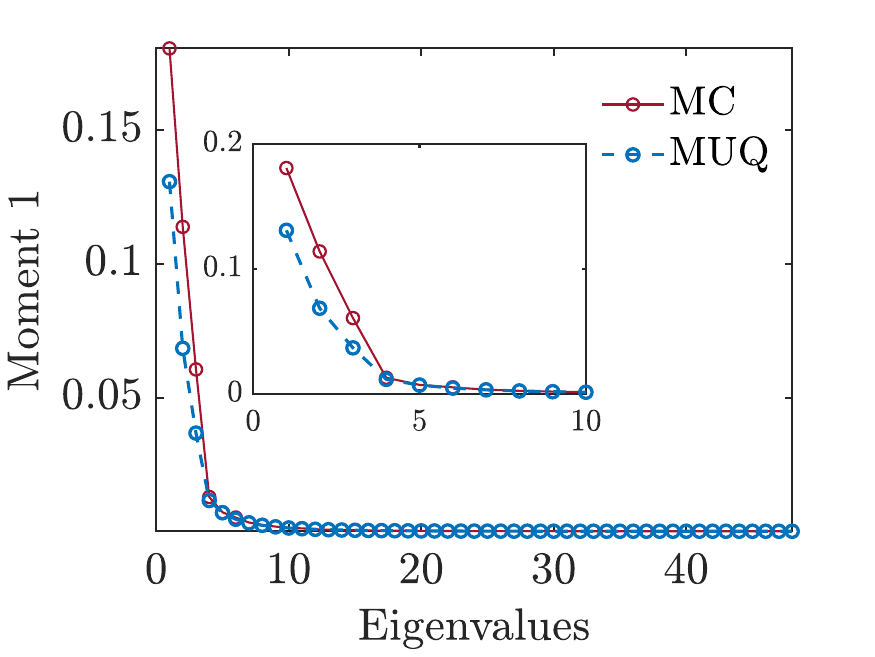} }}%
\subfloat[\centering ]{{\includegraphics[height=4cm,width=4.5cm]{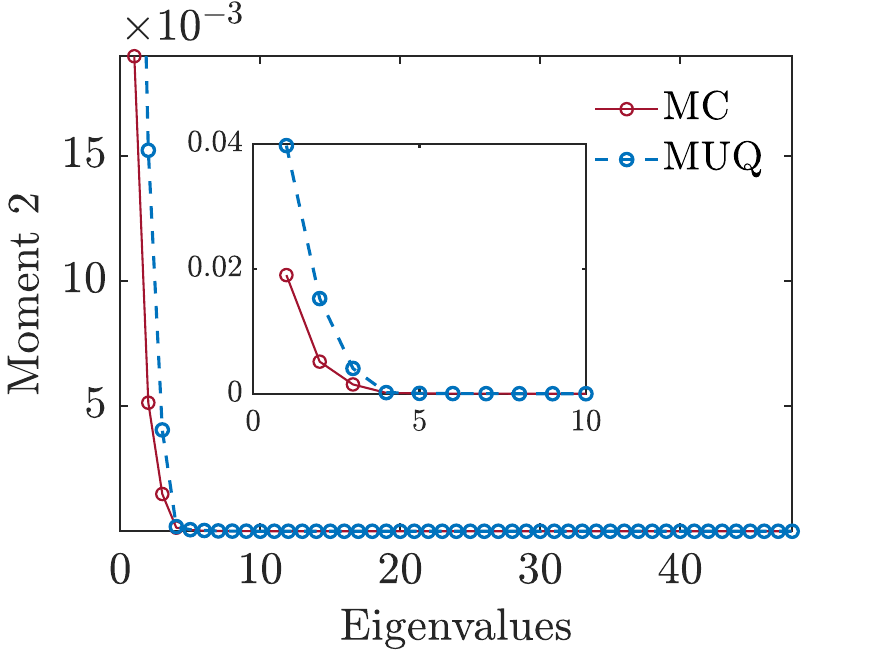} }}%
\quad
\subfloat[\centering ]{{\includegraphics[height=4cm,width=4.5cm]{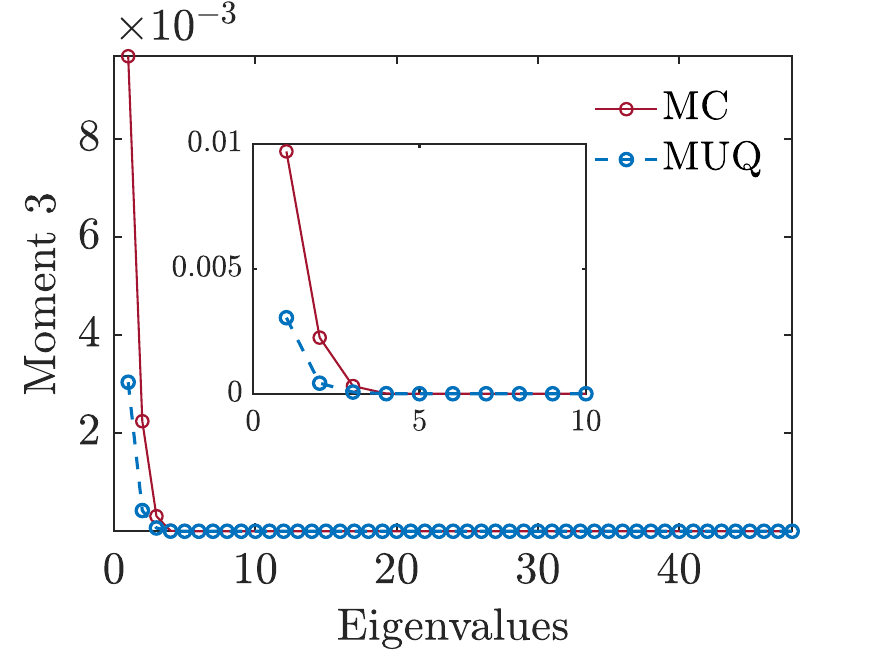} }}%
\subfloat[\centering ]{{\includegraphics[height=4cm,width=4.5cm]{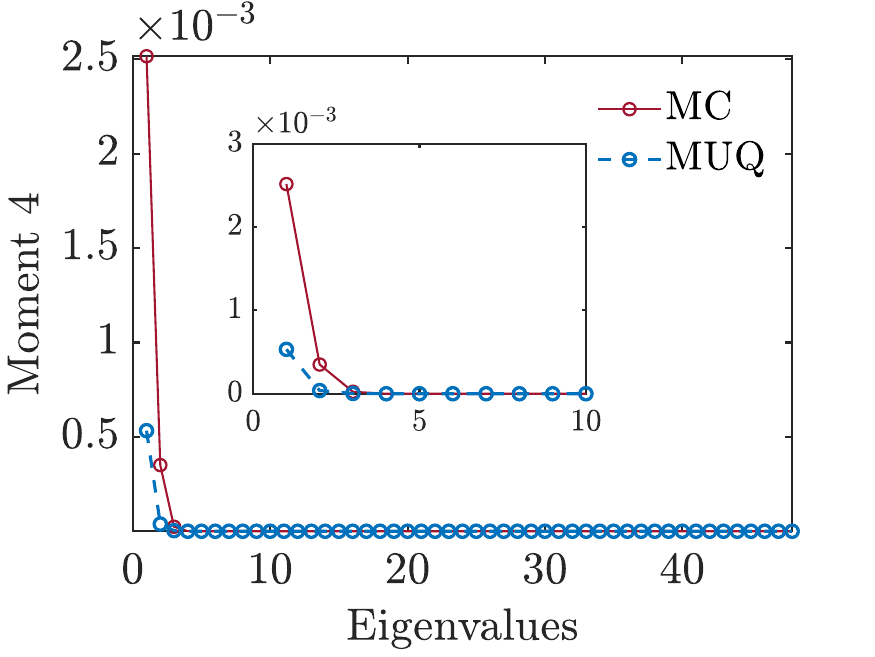} }}%
\caption{Comparison of the first four moments of the $n=48$ Koopman eigenvalues obtained from Monte Carlo and the proposed MUQ.}
    \label{moments_rw}
\end{figure}
\begin{equation}
    K_{ij}=\sum_{k} V_{ki}\lambda_{k}V_{kj};
\end{equation}
\section{Analysis on real sensor measurements}\label{section_realdata}
\subsection{System setup}
The real-world data comprises phasor measurement unit measurements that capture the voltage magnitude and phase angles of a hardware-in-loop system (NREL ARIES) as shown in Figure \ref{fig_ARIES_grid}. Recorded at a sampling frequency of $100$ samples per second, the data span a duration of 10 min. Figure \ref{data_rw1} (a) illustrates one of the steady state states observed. 

To characterize the measurement uncertainty associated with each state, we employ a $9^{th}$-order polynomial regression to model the measured data. Subtracting the regressed values (see Figure \ref{data_rw1} (b)) from the measured data allows us to compute the variance of the difference, serving as the variance estimate for the state. 

\subsection{Results on the Distributions of the Koopman Operator and its Spectral Properties} 
Figure \ref{rw} presents a comparison between the true variance values obtained from the Monte Carlo simulation and those derived from our proposed MUQ method that employs the DMD and EDMD estimation algorithms.
The dimensions of the Koopman operator for the real-world system via the EDMD estimation algorithm for the quadratic polynomial lifting increase to $n=48$.
We plot the spread of the first three elements of the Koopman eigenfunctions along the three dimensions containing the $N$ random realizations of the first four eigenfunctions to better visualize their spread (see Figure \ref{dimensions}).

\begin{figure}[t!]%
\centering
\subfloat[\centering ]{{\includegraphics[height=4cm,width=4.5cm]{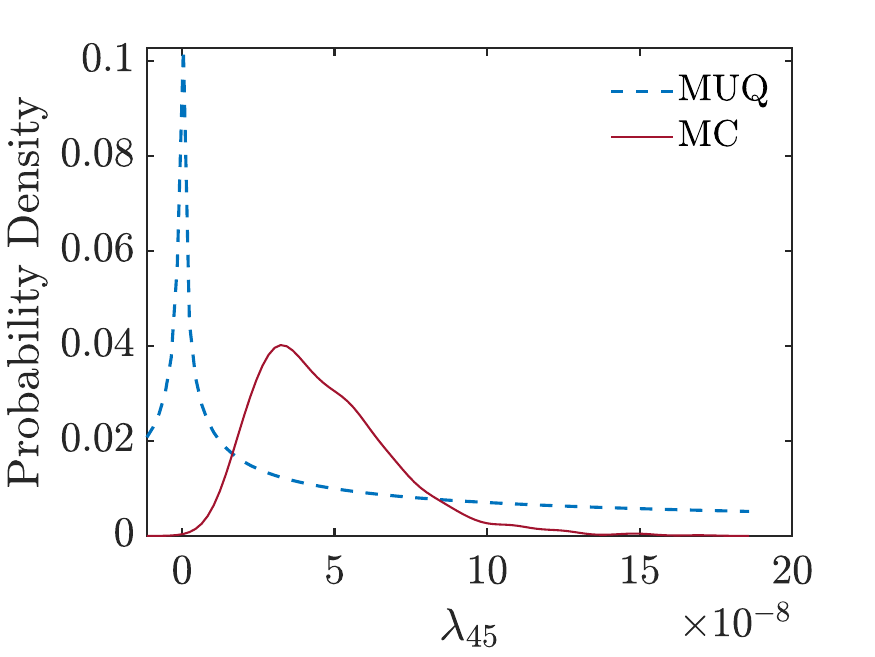} }}%
\subfloat[\centering ]{{\includegraphics[height=4cm,width=4.5cm]{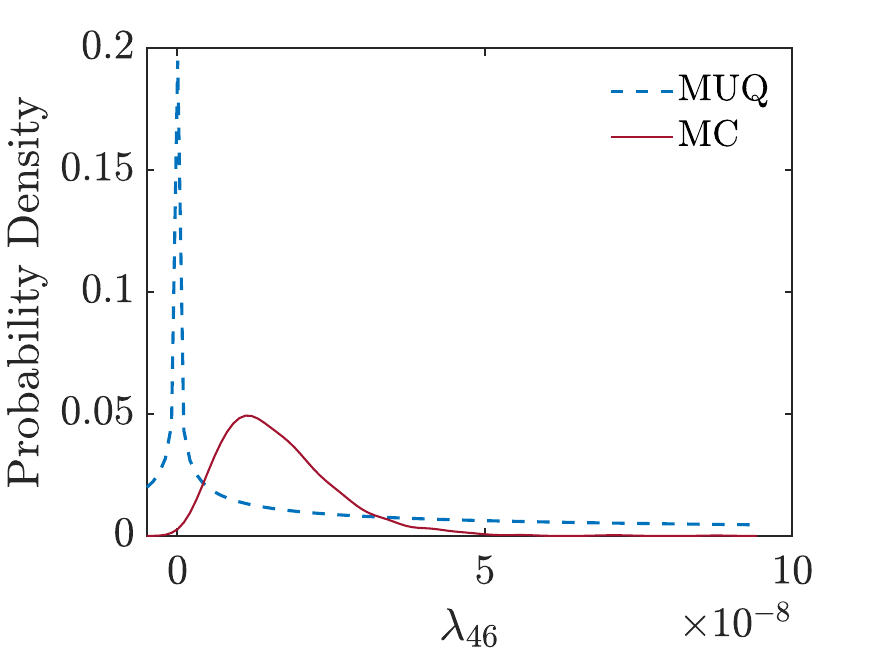} }}%
\quad
\subfloat[\centering ]{{\includegraphics[height=4cm,width=4.5cm]{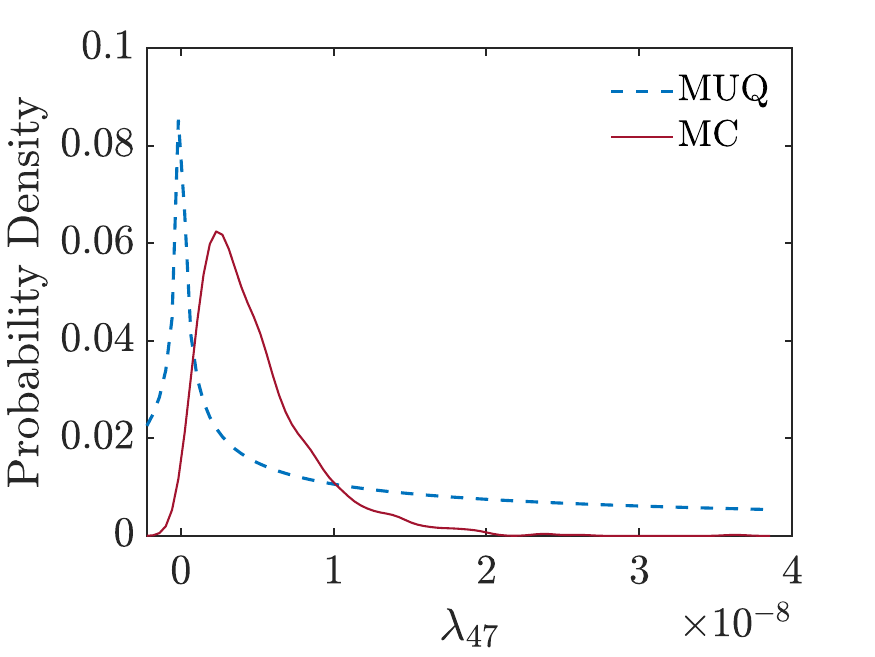} }}%
\subfloat[\centering ]{{\includegraphics[height=4cm,width=4.5cm]{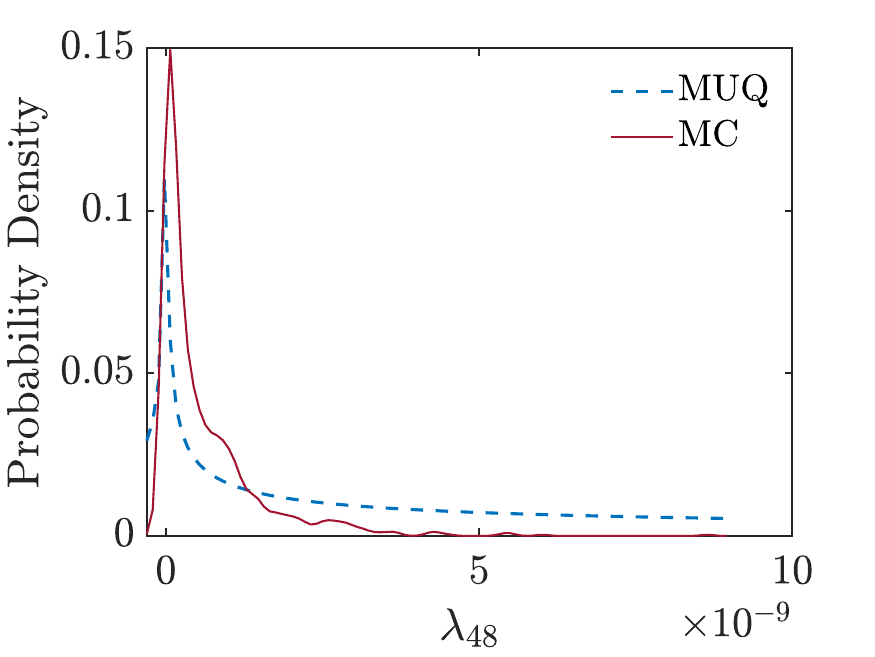} }}%
\caption{Comparison between the probability densities obtained from Monte Carlo and the ones obtained from the proposed uncertainty quantification MUQ of the Koopman eigenvalues (a) $\lambda_{46}$, (b) $\lambda_{47}$, (c) $\lambda_{48}$, and (d) $\lambda_{49}$ for the real-world system. .}%
    \label{4_eigval}
\end{figure}%

We compute the first four moments of the eigenvalues of the estimated Koopman operator. For the Monte Carlo simulation, the $n=48$ eigenvalues are simply the eigenvalues of the $N=1000$ random realizations of the Koopman operator $\bm{K}$ in descending order. A kernel density function is fitted to each eigenvalue considered as random variables to obtain the probability density measure. Random realizations of the Koopman eigenvalues are obtained from the proposed MUQ by using the density function in \eqref{mp} for the Marchenko-Pastur distribution. 
The first four moments of the eigenvalues are calculated from the probability density measure, $m^{(i)}=\int_{-\infty}^{\infty}\lambda^{i}f( \lambda)d\lambda.$ The moments are displayed in Figure \ref{moments_rw}.

The kernel density fit obtained from the proposed method is compared with that obtained from the Monte Carlo method for randomly sampled Koopman eigenvalues, which are ordered $\lambda_{1}\geq \lambda_{2}\hdots\geq \lambda_{n}$, and is represented in Figure \ref{4_eigval}.

The dispersion of the eigenfunctions of the Koopman operator, employed for dynamical system identification from real-world data, is illustrated in Figure \ref{4_eigvec}. The data set of random samples is derived from the proposed QR factorization-based uncertainty quantification and Monte Carlo simulation. Observing the results, we discern a uniform distribution of Koopman eigenfunctions from $+5$ to $-5$, characterized by the Haar measure, $O(n=24)$. The Haar measure ensures uniformity and orthogonality in the spread of Koopman eigenfunctions, contributing to an effective quantification of measurement uncertainty.

Turning our attention to the Koopman eigenvalues, the insights gleaned from Figures \ref{4_eigval_sm} and \ref{4_eigval} affirm that our proposed MUQ method effectively captures the probability distributions of random eigenvalues, ${\lambda}_{1}\geq\lambda_{2}\geq\lambda_{2n}$, within the Koopman operator, $\bm{K}$. In particular, these distributions closely align with those derived from Monte Carlo simulations, validating the accuracy of our proposed approach.

\section{Conclusion}\label{sec5}
We have introduced a methodology, MUQ, that allows the quantification of the uncertainty of the measurements in the Koopman operator based on random matrix theory. This algorithm is straightforward to implement and has been evaluated for its performance using the single-machine three-bus system and on real-world data obtained from hardware in loop system at NREL ARISE. In both cases, our results were compared with values obtained using Monte Carlo simulations.

In our future work, we plan to extend this approach to quantify the uncertainty the tuples of the Koopman operator. We aim to develop analytical expressions for the variances of these tuples in relation to the characterized variances of the states, which represent the measurement uncertainty. In addition, we are committed to improving the accuracy of the proposed MUQ. 
\bibliography{references1} 

\begin{thebibliography}{10}

\bibitem{Wollenberg2012}
A.~J. Wood, B.~F. Wollenberg, and G.~B. Shebl{\'e}, {\em Power generation, operation, and control}.
\newblock John Wiley \& Sons, 2012.

\bibitem{Matevosyan2021}
J.~Matevosyan, J.~MacDowell, N.~Miller, B.~Badrzadeh, D.~Ramasubramanian, A.~Isaacs, R.~Quint, E.~Quitmann, R.~Pfeiffer, H.~Urdal, T.~Prevost, V.~Vittal, D.~Woodford, S.~H. Huang, and J.~O’Sullivan, ``A future with inverter-based resources: Finding strength from traditional weakness,'' {\em IEEE Power Energy Mag.}, vol.~19, no.~6, pp.~18--28, 2021.

\bibitem{fan2022data}
L.~Fan, Z.~Miao, S.~Shah, P.~Koralewicz, V.~Gevorgian, and J.~Fu, ``Data-driven dynamic modeling in power systems: A fresh look on inverter-based resource modeling,'' {\em IEEE Power Energy Mag.}, vol.~20, no.~3, pp.~64--76, 2022.

\bibitem{tan2016survey}
S.~Tan, D.~De, W.-Z. Song, J.~Yang, and S.~K. Das, ``Survey of security advances in smart grid: A data driven approach,'' {\em IEEE Commun. Surv. Tutor.}, vol.~19, no.~1, pp.~397--422, 2016.

\bibitem{IEC61869-9}
``{Instrument transformers - Part 9: Digital interface for instrument transformers},'' {IEC 61869-9}, 2016.

\bibitem{brahma2016real}
S.~Brahma, R.~Kavasseri, H.~Cao, N.~Chaudhuri, T.~Alexopoulos, and Y.~Cui, ``Real-time identification of dynamic events in power systems using {PMU} data, and potential applications—models, promises, and challenges,'' {\em IEEE Trans. Power Deliv.}, vol.~32, no.~1, pp.~294--301, 2016.

\bibitem{farantatos2009pmu}
E.~Farantatos, G.~K. Stefopoulos, G.~J. Cokkinides, and A.~Meliopoulos, ``{PMU}-based dynamic state estimation for electric power systems,'' in {\em IEEE Power \& Energy Society General Meeting}, pp.~1--8, 2009.

\bibitem{Netto2018}
M.~Netto and L.~Mili, ``Robust data filtering for estimating electromechanical modes of oscillation via the multichannel {P}rony method,'' {\em IEEE Trans. Power Syst.}, vol.~33, no.~4, pp.~4134--4143, 2018.

\bibitem{Crow2005}
M.~Crow and A.~Singh, ``The matrix pencil for power system modal extraction,'' {\em IEEE Trans. Power Syst.}, vol.~20, no.~1, pp.~501--502, 2005.

\bibitem{Zhou2006}
N.~Zhou, J.~Pierre, and J.~Hauer, ``Initial results in power system identification from injected probing signals using a subspace method,'' {\em IEEE Trans. Power Syst.}, vol.~21, no.~3, pp.~1296--1302, 2006.

\bibitem{Mauroy2020}
A.~{Mauroy}, I.~{Mezi{\'c}}, and Y.~{Susuki (Editors)}, {\em The Koopman Operator in Systems and Control: Concepts, Methodologies, and Applications}.
\newblock Cham, Switzerland: Springer Nature Switzerland AG, 2020.

\bibitem{Brunton2022}
S.~L. Brunton, M.~Budi\v{s}i\'{c}, E.~Kaiser, and J.~N. Kutz, ``Modern koopman theory for dynamical systems,'' {\em SIAM Review}, vol.~64, no.~2, pp.~229--340, 2022.

\bibitem{susuki2011nonlinear}
Y.~Susuki and I.~Mezi{\'c}, ``Nonlinear {K}oopman modes and coherency identification of coupled swing dynamics,'' {\em IEEE Trans. Power Syst.}, vol.~26, no.~4, pp.~1894--1904, 2011.

\bibitem{susuki2013nonlinear}
Y.~Susuki and I.~Mezi{\'c}, ``Nonlinear {K}oopman modes and power system stability assessment without models,'' {\em IEEE Trans. Power Syst.}, vol.~29, no.~2, pp.~899--907, 2013.

\bibitem{Netto2018AEstimation}
M.~Netto and L.~Mili, ``A robust data-driven {K}oopman {K}alman filter for power systems dynamic state estimation,'' {\em IEEE Trans. Power Syst.}, vol.~33, no.~6, pp.~7228--7237, 2018.

\bibitem{susuki2018estimation}
Y.~Susuki, R.~Hamasaki, and A.~Ishigame, ``Estimation of power system inertia using nonlinear {K}oopman modes,'' in {\em IEEE Power \& Energy Society General Meeting}, pp.~1--5, 2018.

\bibitem{Netto2019}
M.~Netto, Y.~Susuki, and L.~Mili, ``Data-driven participation factors for nonlinear systems based on {K}oopman mode decomposition,'' {\em IEEE Contr. Syst. Lett.}, vol.~3, no.~1, pp.~198--203, 2019.

\bibitem{Sharma2022Data-DrivenMeasurements}
P.~Sharma, V.~Ajjarapu, and U.~Vaidya, ``Data-driven identification of nonlinear power system dynamics using output-only measurements,'' {\em IEEE Trans. Power Syst.}, vol.~37, no.~5, pp.~3458--3468, 2022.

\bibitem{Sinha2020DataDynamics}
S.~Sinha, S.~P. Nandanoori, and E.~Yeung, ``Data driven online learning of power system dynamics,'' in {\em IEEE Power \& Energy Society General Meeting}, pp.~1--5, 2020.

\bibitem{Schmid2010}
P.~J. Schmid, ``Dynamic mode decomposition of numerical and experimental data,'' {\em J. Fluid Mech.}, vol.~656, p.~5–28, 2010.

\bibitem{Rowley2009}
C.~W. Rowley, I.~Mezi{\'c}, S.~Bagheri, P.~Schlatter, and D.~S. Henningson, ``Spectral analysis of nonlinear flows,'' {\em J. Fluid Mech.}, vol.~641, p.~115–127, 2009.

\bibitem{Barocio2015}
E.~Barocio, B.~C. Pal, N.~F. Thornhill, and A.~R. Messina, ``A dynamic mode decomposition framework for global power system oscillation analysis,'' {\em IEEE Trans. Power Syst.}, vol.~30, no.~6, pp.~2902--2912, 2015.

\bibitem{duke2012error}
D.~Duke, J.~Soria, and D.~Honnery, ``An error analysis of the dynamic mode decomposition,'' {\em Exp. Fluids}, vol.~52, pp.~529--542, 2012.

\bibitem{Dawson2014CharacterizingDecomposition}
S.~T.~M. Dawson, M.~S. Hemati, M.~O. Williams, and C.~W. Rowley, ``{Characterizing and correcting for the effect of sensor noise in the dynamic mode decomposition},'' {\em Exp. Fluids}, vol.~57, 2014.

\bibitem{nuske2023finite}
F.~N{\"u}ske, S.~Peitz, F.~Philipp, M.~Schaller, and K.~Worthmann, ``Finite-data error bounds for {K}oopman-based prediction and control,'' {\em J. Nonlinear Sci.}, vol.~33, no.~1, p.~14, 2023.

\bibitem{Williams2015ADecomposition}
M.~O. Williams, I.~G. Kevrekidis, and C.~W. Rowley, ``A data–driven approximation of the {K}oopman operator: Extending dynamic mode decomposition,'' {\em J. Nonlinear Sci.}, vol.~25, no.~6, pp.~1307--1346, 2015.

\bibitem{zhang2023quantitative}
C.~Zhang and E.~Zuazua, ``A quantitative analysis of koopman operator methods for system identification and predictions,'' {\em Comptes Rendus. M{\'e}canique}, vol.~351, no.~S1, pp.~1--31, 2023.

\bibitem{Lian2020OnOperators}
Y.~Lian and C.~N. Jones, ``On {G}aussian process based {K}oopman operators,'' {\em IFAC-PapersOnLine}, vol.~53, no.~2, pp.~449--455, 2020.

\bibitem{Xu2022PropagatingModel}
Y.~Xu, M.~Netto, and L.~Mili, ``Propagating parameter uncertainty in power system nonlinear dynamic simulations using a {K}oopman operator-based surrogate model,'' {\em IEEE Trans. Power Syst.}, vol.~37, no.~4, pp.~3157--3160, 2022.

\bibitem{Matavalam2022PropagatingOperators}
A.~R.~R. Matavalam, U.~Vaidya, and V.~Ajjarapu, ``Propagating uncertainty in power system initial conditions using data-driven linear operators,'' {\em IEEE Trans. Power Syst.}, vol.~37, no.~5, pp.~4125--4128, 2022.

\bibitem{Cook2011OnMatrix}
R.~D. Cook and L.~Forzani, ``On the mean and variance of the generalized inverse of a singular {W}ishart matrix,'' {\em Electron. J. Stat.}, vol.~5, pp.~146--158, 2011.

\bibitem{Lopuhaa1991BreakdownMatrices}
H.~P. Lopuhaa and P.~J. Rousseeuw, ``Breakdown points of affine equivariant estimators of multivariate location and covariance matrices,'' {\em Ann. Stat.}, vol.~19, no.~1, pp.~229--248, 1991.

\bibitem{Marcenko1967DistributionMatrices}
V.~A. Mar{\v{c}}enko and L.~A. Pastur, ``Distribution of eigenvalues for some sets of random matrices,'' {\em Mathematics of the USSR-Sbornik}, vol.~1, no.~4, pp.~457--483, 1967.

\bibitem{Bai2007OnMatrix}
Z.~D. Bai, B.~Q. Miao, and G.~M. Pan, ``On asymptotics of eigenvectors of large sample covariance matrix,'' {\em Ann. Probab.}, vol.~35, no.~4, pp.~1532--1572, 2007.

\bibitem{Silverstein1981DescribingGroups}
J.~W. Silverstein, ``Describing the behavior of eigenvectors of random matrices using sequences of measures on orthogonal groups,'' {\em SIAM J. Math. Anal.}, vol.~12, no.~2, pp.~274--281, 1981.

\bibitem{dobson_model}
I.~Dobson, H.-D. Chiang, J.~Thorp, and L.~Fekih-Ahmed, ``A model of voltage collapse in electric power systems,'' in {\em IEEE Conf. Decis. Control}, vol.~3, pp.~2104--2109, 1988.

\bibitem{ajjarapu_bifurcation}
V.~Ajjarapu and B.~Lee, ``Bifurcation theory and its application to nonlinear dynamical phenomena in an electrical power system,'' {\em IEEE Trans. Power Syst.}, vol.~7, no.~1, pp.~424--431, 1992.

\end{thebibliography}
\bibliographystyle{ieeetr}

\end{document}